\documentclass[10pt,a4paper]{article}
\usepackage{geometry}
\usepackage{amsfonts}
\usepackage{amsmath}
\usepackage{amssymb}
\usepackage{comment}
\usepackage{mathrsfs}
\usepackage{graphicx}
\usepackage{hyperref}
\usepackage[usenames,dvipsnames]{color}
\usepackage{cancel}
\usepackage{cleveref}
\usepackage{bbm}
\usepackage{bm}
\usepackage{mathtools}
\usepackage{physics}
\usepackage{tensor}
\usepackage{array}
\usepackage{float}
\usepackage{wrapfig}
\usepackage{soul}
\usepackage{xcolor}
\usepackage{cite}
\usepackage{url}
\usepackage{multicol}

\usepackage{tocloft}

\newcommand{\flip}{\sigma}

\hypersetup{
colorlinks=true,         
citecolor=blue,           
}

\title{\vspace{-2cm} \huge Multiparticle states in braided lightlike\\
 $\kappa$-Minkowski noncommutative QFT}
\date{}
\author{{$^{(1)}$Giuseppe Fabiano\footnote{peppefabiano@hotmail.com}, $^{(2)}$Flavio Mercati\footnote{flavio.mercati@gmail.com}}
\vspace{12pt}
\\
\small  $^{(1)}$\small Dipartimento di Fisica Ettore Pancini, Universit\`a di Napoli ``Federico II'';
\\
\small and INFN, Sezione di Napoli, Complesso Univ. Monte S. Angelo, I-80126 Napoli, Italy;\\
\small $^{(2)}$Departamento de F\'isica, Universidad de Burgos, 09001 Burgos, Spain.
}

\begin{document}

\maketitle

\begin{abstract}
In this study, we construct a 1+1-dimensional, relativistic, free, complex scalar Quantum Field Theory on the noncommutative spacetime known as lightlike  $\kappa$-Minkowski. The associated $\kappa$-Poincar\'e quantum group of isometries is triangular, and its quantum R matrix enables the definition of a braided algebra of N points that retains $\kappa$-Poincaré invariance.
Leveraging our recent findings, we can now represent the generators of the deformed oscillator algebra as nonlinear redefinitions of undeformed oscillators, which are nonlocal in momentum space.
The deformations manifest at the multiparticle level, as the one-particle states are identical to the undeformed ones.
We successfully introduce a covariant and involutive deformed flip operator using the R matrix. The corresponding deformed (anti-)symmetrization operators are covariant and idempotent, allowing for a well-posed definition of multiparticle states, a result long sought  in Quantum Field Theory on $\kappa$-Minkowski. We find that P and T are not symmetries of the theory, although PT (and hence CPT) is. We conclude by noticing that identical particles appear distinguishable in the new theory, and discuss the fate of the  Pauli exclusion principle in this setting.
\end{abstract}

\vspace*{\fill}

 \tableofcontents

\vspace*{\fill}

\newpage

\section{Introduction}

Quantum Field Theory (QFT) on noncommutative spacetimes has been studied for decades~\cite{Szabo:2001kg,aschieri2011noncommutative,Wulkenhaar2019}. The initial motivations came from the desire to regularize the ultraviolet divergences of quantum electrodynamics~\cite{Snyder1947}, together with the intuition that quantum theory might require spacetime itself to be quantized in some sense. This proposal entails some serious interpretational challenges, as the smooth topology of spacetime would have to be replaced by something new. The discussion of these issues echoes the debate that followed the introduction of quantization conditions on phase-space orbits by Bohr~\cite{GamowBook}. The idea of describing quantum phase space as a non-smooth, ``pointless'' geometry remained in the back burner until the development of the theory of von Neumann algebras, which represented the birth of noncommutative geometry~\cite{ConnesBook}. The commutative $C^*$-algebras of functions on a topological manifold have been shown to completely characterize the topology of the manifold~\cite{gelfand1943imbedding,segal1947irreducible}. Replacing this algebra with a noncommutative algebra leads to the modern notion of a noncommutative geometry~\cite{ConnesBook}. The tools of classical differential topology and Riemannian geometry are insufficient to describe objects that lack the notion of infinitesimal points,  so the study of the properties of these noncommutative spaces requires a purely algebraic formulation. Connes, Woronowicz and Drinfel'd  generalized the notion of a differential structure to the noncommutative setting~\cite{Connes1985,Woronowicz1987_1,Woronowicz1987_2}, 
which led to the definition of gauge theories on a large class of noncommutative spaces. The ``Noncommutative Standard Model'' of Connes, Lott and Chamseddine~\cite{connes1989particle,chamseddine1997spectral} allows to unify the Standard Model fields with the Higgs boson as well as the gravitational field as an effective description of a ``quasi-commutative'' geometry. This model has been extensively studied for decades, and, although so far it has not been possible to deduce unambiguous predictions of some Standard Model parameters that would make the model falsifiable, it still represents today a possible avenue towards unification.

The main motivation for a noncommutative structure of spacetime comes from Quantum Gravity.  Already at the effective level of QFT coupled to classical general relativity, it can be shown that the Planck length, $\ell_P=\sqrt{{G\hbar}/{c^3}}$, represents a limit to the localizability of fields~\cite{Hossenfelder:2012jw}. Rather than attacking frontally the full problem of Quantum Gravity, one could incorporate these restrictions to locality in effective models that feature uncertainty relations among some noncommutative ``coordinate'' operators. A compelling evidence that such a ``noncommutative QFT'' could be a realistic, intermediate level of description between commutative QFT and the full quantum theory of gravity comes from the only model of Quantum Gravity that is well-understood as a QFT: 2+1-dimensional General Relativity. This theory lacks local propagating degrees of freedom (gravitons) and can be therefore quantized with topological QFT methods. Coupling the theory to matter and integrating away the gravitational degrees of freedom gives rise to an effective theory of matter propagating on a noncommutative spacetime~\cite{Matschull:1997du,Freidel:2005me}. The Planck scale in this model ends up playing the role of a scale of noncommutativity.

Further evidence for spacetime noncommutativity as an effective description comes from String Theory, in which the  intrinsic length scale of strings has been conjectured to prevent probing arbitrarily small distances~\cite{veneziano1986stringy}. Moreover, a connection between String Theory and noncommutative geometry was found by Witten in the context of interacting bosonic open strings~\cite{witten1986noncommutative}, and later with Seiberg, with the identification of a regime in which the string dynamics is described by a gauge theory on a noncommutative space~\cite{Seiberg:1999vs}. Models of noncommutative QFT inspired by the Seiberg--Witten one (so-called $\theta$-Moyal-type spacetimes) have been extensively studied in their own right~~\cite{Szabo:2001kg}, and a 1+1-dimensional toy model of this kind, studied by Grosse and Wulkenhaar~\cite{Grosse:2003nw}, was proven to be finite at all energy. This is the only known example of an interacting QFT that is well-defined at all scales, and realizes Snyder's original dream of keeping the divergences of QFT under control via noncommutativity.
Notice that the Seiberg--Witten model~\cite{Seiberg:1999vs} relies on a tensor B-field taking an expectation value in the vacuum and providing a Lorentz-breaking preferred background frame with respect to  which the noncommutativity of coordinates is specified.
This, of course, raises the question of the destiny of Lorentz invariance in noncommutative QFT, which becomes particularly pressing in view of the stringent constraints available today on Lorentz invariance violations \cite{Mattingly:2005re,Bolmont:2022yad}. Beyond the need to evade these constraints, we would also like to establish whether spacetime noncommutativity can be compatible with relativistic invariance, which is a question of great interest in itself. That the minimal lengths appearing in Quantum Gravity might not imply Lorentz invariance violation has been long conjectured. In Loop Quantum Gravity, it has been argued that the discreteness of the spectra of geometric operators like the area is analogue to what happens in quantum mechanics with the angular momentum operator~\cite{Rovelli_2003}: one can have a rotationally-symmetric state of non-zero spin, which of course is impossible in classical mechanics. In quantum mechanics, these rotationally-invariant states appear the same to all observers, because, although upon measurement they get nonzero spins along one direction, this happens with a spherically-symmetric probability distribution, and the symmetry is only broken upon choosing a direction in space and realizing a projective measurement.

In this paper, we are interested in conjectured field theories that \textit{deform,} rather than \textit{break,} Poincar\'e invariance. In a theory of this type, the scale of noncommutativity would represent a second relativistic scale, on par with the speed of light $c$, which appears the same to all observers. This would be a concrete realization of the principle of \textit{Doubly Special Relativity} conjectured more than two decades ago by Amelino-Camelia~\cite{Amelino-Camelia:2000stu}. In such a theory, the transformation rules between inertial observers are changed in such a way that not only the speed of light, but also a new length/energy scale appears the same to all observers.
The existence of deformed-Poincar\'e-invariant QFTs has long been conjectured, and models that may qualify as such have been studied extensively for decades, encountering severe conceptual and technical difficulties. The main reason to believe that such theories exist comes again from noncommutative geometry: there are many examples of \textit{noncommutative homogeneous spaces,} which are invariant under \textit{quantum group} symmetries~\cite{majid_1995,majid_2002}. These are generalizations of the notion of a Lie group, in which the algebra of functions on the group is noncommutative, and the transformation parameters become fuzzy, just like the coordinates of the space. Deforming the Poincaré group into a quantum group depending on an invariant length scale seems therefore to be a concrete realization of the Doubly Special Relativity principle mentioned above~\cite{Majid:1988we}.
We now have a few candidates for  a \textit{noncommutative Minkowski spacetime,} invariant under a \textit{quantum Poincar\'e group}~\cite{Gracia-Bondia:2006nhx, Majid:1988we,Fabiano:2023uhg}, however the debate on how to build a QFT on such spaces in such a way that Poincar\'e invariance is preserved is still open. 
In the case of $\theta$-Moyal-type noncommutative spacetimes, the standard techniques used to derive QFT predictions from the theory is to introduce a \textit{star product,} which is an infinite-dimensional representation of the noncommutative product between coordinates on a space of commutative functions. Then the noncommutative action of a field theory can be written as a nonlocal action in terms of commutative fields. The nonlocality is due to the fact that the action depends on derivatives of all orders of the commutative fields. Treating such an action as a commutative one, one gets correlation functions that depend on the noncommutativity parameters in an apparently Lorentz-violating way. This seems at odds with the existence of quantum group deformations of the Poincar\'e group that leave the $\theta$-Moyal-type noncommutative spacetime (and the corresponding QFT actions) invariant~\cite{chaichian2004lorentz,WESS_2005,douglas2001noncommutative,oeckl2000untwisting}. Solutions have been proposed, that require a distinction between active and passive transformations~\cite{Bichl_2002,matlock2005noncommutative,Vitale:2023znb}. In~\cite{Fiore:2007vg,Fiore:2008ta}, a different approach was proposed, that involves a more careful treatment of the concept of \textit{multilocal functions.} In the commutative case, these are simply functions from several points on the spacetime manifold onto the complex or real numbers, and in terms of the (commutative) algebra of functions on the manifold, they can be simply formulated as elements of tensor products of copies of the same algebra of functions. In the noncommutative setting, taking simply the tensor product fails to produce a covariant structure: in other words, assuming that the coordinates of different points commute with each other is not covariant under the quantum group Poincar\'e transformations. Fortunately, mathematicians found a generalization of the concept of tensor product, called \textit{braiding}~\cite{majid_1995}, which allows one to identify a noncommutative algebra of N-points that is invariant under the relevant quantum group. In~\cite{Fiore:2007vg,Fiore:2008ta}, this structure was used to define the Wightman functions of a noncommutative QFT in a consistent way. Interestingly, the difference between noncommutative coordinates of different points belong to a commutative subalgebra of the braided N-point algebra, so that the N-point functions of the theory are commutative and admit a simple interpretation as correlation functions in perfect analogy to commutative QFT. There have been various attempts to produce testable predictions of QFT on the $\theta$-Moyal spacetime, including some conjectured violations of the Pauli Exclusion principle~\cite{Balachandran:2005eb,Addazi:2017bbg,Addazi:2018ioz}. The original approach to $\theta$-Moyal braided QFT suggests that the theory is equivalent to the commutative one, at least at the perturbative level~\cite{Fiore:2007zz}. More recent approaches \cite{Nguyen:2021rsa,DimitrijevicCiric:2021jea,Giotopoulos:2021ieg,DimitrijevicCiric:2023hua,Bogdanovic:2023izt}, based on Oeckl's algebraic definition of braided QFT~\cite{Oeckl:1999zu,Oeckl:2000eg}, find that N-point functions for $N>3$ are indeed deformed~\cite{DimitrijevicCiric:2023hua}, although the one-loop self-energy of a $\lambda \, \phi^4$ theory is not. The approach of \cite{Nguyen:2021rsa,DimitrijevicCiric:2021jea,Giotopoulos:2021ieg,DimitrijevicCiric:2023hua,Bogdanovic:2023izt} works for any quantum group deformation of the Poincar\'e group that can be expressed as a twist, and therefore applies, in principle, to the case studied in the present paper. It would be interesting to compare that approach with ours.

Recently, some of us developed a similar construction for QFT on a different noncommutative spacetime~\cite{Lizzi:2021rlb,DiLuca:2022idu}, finding a rich and previously unnoticed level of complexity. The noncommutative geometry in question is the so-called ``lightlike'' $\kappa$-Minkowski spacetime, symmetric under the ``lightlike'' $\kappa$-Poincar\'e quantum group~\cite{Ballesteros:1993zi,Ballesteros:1995mi,Ballesteros1997,Ballesteros:1996awf}. This model has great interest for physics, because its noncommutativity parameter has the dimensions of a length (unlike the $\theta$-Moyal models whose parameter is an \textit{area}), which suggests that it might capture the first order in an expansion in powers of the Planck length of an effective theory of fields on a quantum-gravitational background. 
A closely related noncommutative spacetime (the so-called ``timelike'' version of $\kappa$-Minkowski) has been studied for decades~\cite{Lukierski:1991ff,Lukierski:1991pn,Lukierski:1992dt,Majid:1994cy,Majid:1999td,Lukierski:2015zqa,Lizzi:2018qaf,Lizzi:2019wto,Lukierski:1993wxa,Kowalski-Glikman:2002eyl,Agostini:2002yd,Agostini:2002de,Kowalski-Glikman:2002oyi,Kowalski-Glikman:2003qjp,Agostini:2005mf,Amelino-Camelia:2007yca,Amelino-Camelia:2007rym,Carmona:2011wc,Carmona:2012un,Amelino-Camelia:2010yrn,Gubitosi:2011hgc,Mercati:2011aa,Meljanac:2016jwk,Loret:2016jrg,Lukierski:2016vah,Mercati:2018fba}, and many efforts went into building a consistent QFT on it~\cite{Kosinski:1999ix,Kosinski:2001ii,Kosinski:2003xx,Arzano:2007ef,Daszkiewicz:2007az,Freidel:2007hk,Arzano:2009ci,Arzano:2017uuh,Juric:2015hda,Mathieu:2020ccc,Poulain:2018mcm,Poulain:2018two,Juric:2018qdi,Arzano:2018gii,Mercati:2018hlc,Mercati:2018ruw}. A lot of progress has been made on the problem, although much work remains to be done, and a fully satisfactory QFT, both from the interpretational and the technical/mathematical point of view, remains elusive.

In our first paper~\cite{Lizzi:2021rlb}, the covariant braided N-point algebra for a general parametrization of $\kappa$-Minkowski-like noncommutative spacetimes was constructed, and it was proven that its associativity is only compatible with the lightlike model. In fact,  the structure constants of the coordinate algebra of this class of spacetimes are usually expressed in terms of four parameters forming a vector (conventionally called $v^\mu$), and the braiding construction turned out to be possible only if said vector is lightlike (or null), hence the name of the model. Furthermore, the coordinate differences between different points (and therefore all N-point functions) were shown to be commutative, just like in the work of Wess and Fiore~\cite{Fiore:2007vg,Fiore:2008ta}. In~\cite{Lizzi:2021rlb}, a proposal for a covariant Pauli--Jordan function was put forward, however a technical obstacle prevented us from defining general Lorentz-invariant N-point functions. Namely, the momentum space of the theory was not closed under Lorentz transformations, which practically meant that certain momentum space integrals would have a Lorentz-breaking upper bound related to the deformation energy scale. Thanks to a recent observation~\cite{Bevilacqua:2022fbz}, this problem of the non-closure of momentum space under Lorentz transformations can be solved by enlarging the basis of noncommutative functions that are used in the Fourier expansion of fields, to plane waves that include a constant complex contribution to the frequency. This allows one to ``double'' momentum space into two halves that are connected to each other by Lorentz transformations, and together, are globally Lorentz invariant. This observation was used in the recent work~\cite{DiLuca:2022idu} to define a free complex scalar field theory consistently (using covariant quantization based on a Pauli--Jordan function), and derive the associated deformed construction and annihilation operator algebra. Unfortunately, this algebra turned out to be extremely complicated, due to the presence of the additional region of momentum space, which, together with the mass shell, splits the commutator of two creation/annihilation operators into no less than \textit{twenty} cases which need to be listed separately.

In this paper, we build upon the results of~\cite{Lizzi:2021rlb} and~\cite{DiLuca:2022idu}, and find a substantial simplification for the oscillator algebra. We are able to find a simple representation for our deformed creation and annihilation operators that can be expressed in one line, and is based on infinite nonlinear combinations of standard creation and annihilation operators. Such representation makes the unwieldy algebra of~\cite{DiLuca:2022idu} treatable, and allows us to begin drawing some physical conclusions from the theory. First, the one-particle sector is completely undeformed and coincide with that of a commutative free complex scalar QFT. Secondly, the charge conjugation operator is undeformed and Poincar\'e covariant. This was not the case in other approaches to QFT on $\kappa$-Minkowski (in the case of timelike $v^\mu$)~\cite{Bevilacqua:2022fbz}. In particular, the recent~\cite{Arzano:2020jro} shows that the charge conjugation operator sends a one-particle state into a one-antiparticle state with different momentum. This phenomenon is not present in our model. Regarding P and T symmetries, these are not symmetries of the commutation relations between coordinates, and this fact manifests itself already at the level of the one-particle sector: we are not able to introduce a P or a T operator that acts on the oscillator algebra or on the Fock space in the desired way. This, however, does not prevent PT symmetry from being realized: thanks to the antilinearity of the T operator, both the coordinate commutation relations and the free field theory can be shown to be PT-invariant. Having C and PT, the CPT invariance of the model is also guaranteed.

The nontriviality of the model manifests itself all in the multi-particle sector. Already at the level of two particles one sees that the total momentum depends nonlinearly on the momenta of the two particles, and the action of Lorentz transformations on two momenta becomes nonlinear and mixes the components of the momenta of the two particles (something dubbed ``backreaction'' in previous works~\cite{Gubitosi:2011hgc,Majid:2006xn}). Finally, we are able to introduce a  ``braided flip operator'' that exchanges the momenta of two particles in a nonlinear way, which possesses all the properties that such an operator should: it is Lorentz covariant and is an involution (its square is the identity operator). This operator can be used to define symmetric and antisymmetric states, which are necessary to define the Fock space of bosonic and fermionic fields. Recent work by another group~\cite{Arzano:2022vmh} showed that, in the case of the ``timelike''  $\kappa$-Minkowski spacetime, such a flip operator does not exist. The next best things are either non-Lorentz-covariant at all orders, or are not involutive~\cite{Arzano:2008bt,Arzano:2009wp,Govindarajan:2009wt,Arzano:2013sta}, which means that one can build an infinite tower of two-particle states that all share the same total momentum. The conclusion of the authors of~\cite{Arzano:2022vmh} is that the very notion of identical particles, and (anti-)symmetrized multiparticle states loses meaning. These results do not apply to the model considered in the present paper, as our flip operator is both involutive and Lorentz-covariant. This allows us to introduce a well-defined notion of multi-particle states, which is something that has eluded studies of QFT on $\kappa$-Minkowski for decades. The deformed multi-particle states allow for a revision of the classical concepts of indistinguishability of identical particles and of the Pauli exclusion principle. We find that, given enough precision, particles of the same species which are described by a deformed (anti)-symmetric state can be distinguished by an experiment measuring their momenta. Moreover, the class of states prohibited by the Pauli Exclusion Principle is instead allowed in this setting, while another class of states not excluded by the standard Principle is instead prohibited.

\section{Noncommutative geometry of lightlike $\kappa$-Minkowski}

\subsection{The lightlike $\kappa$-Minkowski spacetime and the $\kappa$-Poincar\'e quantum group}
\label{sec:kspaceandgroup}

The $d+1$-dimensional $\kappa$-Minkowski noncommutative space-time is defined by commutations relations among coordinates of the form
\begin{equation} \label{eq:kmink}
    [x^{\mu},x^{\nu}]=\frac{i}{\kappa}(v^{\mu}x^{\nu}-v^{\nu}x^{\mu}) \,,  \quad \mu = 0,1,\dots,d \,,
\end{equation}
where $\kappa$ is a deformation parameter with the dimensions of energy (in natural units $c=\hbar =1$), and $v^\mu$ is a set of four real parameters. The algebra of functions on Minkowski space-time is hence deformed into a non-commutative algebra $\mathcal{A}$, generated by $x^{\mu}$ and the identity $1$, and equipped with a non-commutative product defined by \eqref{eq:kmink}.
One can introduce a (commutative) arbitrary constant metric tensor $g_{\mu\nu}$, and require that it is preserved by a quantum group of symmetries, which also leaves leaves the commutation relations~\eqref{eq:kmink} invariant. One obtains different quantum groups depending on the relationship between the parameters $v^\mu$ and the metric $g_{\mu\nu}$. If the parameters form a lightlike/null vector, \textit{i.e.} $v^\mu v^\nu g_{\mu\nu} = 0$, one obtains a \textit{triangular} Hopf algebra~\cite{majid_1995}, which is the best-behaved case (see below). This quantum group has been discovered in~\cite{Ballesteros:1993zi,Ballesteros:1995mi,Ballesteros1997,Ballesteros:1996awf}. The spacelike case has been discussed in \cite{Blaut:2003wg,Lizzi:2020tci}, while the timelike one, first introduced in~ \cite{Lukierski:1992dt,Lukierski:1991ff,Zakrzewski_1994},  is by far the most-studied one. The appeal of the timelike case  is that, superficially, the algebra~\eqref{eq:kmink} appears spatially isotropic, and indeed it is invariant under commutative/undeformed spatial rotations. At an early time of investigation of the physics of quantum groups, when some phenomenological consequences were being conjectured, undeformed spatial isotropy seemed compelling, because before clarifying the difference between symmetry breaking and symmetry deformations, a non-isotropic model could be feared to be incompatible with very basic observations of the isotropy of empty space \cite{Amelino-Camelia:2003xax}. At the present stage of understanding of the model, these worries result unfounded. Of course~\eqref{eq:kmink} cannot be invariant under the full Lorentz (or Poincar\'e) group, unless one replaces the group with a quantum group, as we will show momentarily. In this case, whether the commutators~\eqref{eq:kmink} appear spatially isotropic or not is an irrelevant point: the only way this affects the theory is that there is a basis for the quantum Poincar\'e algebra of invariance of~\eqref{eq:kmink} in which the rotation generators appear ``more commutative/undeformed'' (specifically: their \textit{coproducts are primitive}). This does not have any real consequences, because, as we are about to show, there is a sense in which, for any choice of $v^\mu$, the $\kappa$-Poincar\'e group has a Lorentz subgroup that is commutative/undeformed, and all the noncommutativity is relegated to the translations, which act on the Lorentz group in a nontrivial way. For these reasons, we do not find any valid reason to prefer a particular choice of $v^\mu$ vector, at this stage.

In developing field theories on the non-commutative space-time~~\eqref{eq:kmink}, the notion of $N$-point functions is essential. These can be defined starting from the braided tensor product algebra $\mathcal{A}^{\tilde \otimes N}$, which deforms the standard tensor product of $N$ copies of $\mathcal{A}$ by introducing nontrivial commutation relations between the coordinates of different points, like $[x^\mu_a,x^\nu_b]\neq 0$, with $a,b$ referring to different copies of $\mathcal{A}$.
Details of this construction can be found in \cite{Lizzi:2021rlb,DiLuca:2022idu}, where it was also shown that the $\kappa$-Poincaré invariance of the $\mathcal{A}^{\tilde \otimes N}$ commutation relations, together with the imposition of the  Jacobi rule, selects only $v^\mu$ such that $g_{\mu\nu}v^{\mu}v^{\nu}=0$.

Therefore, from now on (and as done in \cite{Lizzi:2021rlb,DiLuca:2022idu}), we will restrict our attention to the lightlike $\kappa$-Minkowski non-commutative space-time, which in $1+1$ dimensions is characterized by the commutation relations
\begin{equation}
\label{eq:cnc}
    [x^+,x^-]=\frac{2i}{\kappa} \, x^- \,, \qquad x^\pm = x^0 \pm x^1 \, .
\end{equation}
In the following, we will work in units in which $\kappa=1$.

The symmetries of \eqref{eq:cnc} are expressed in terms of the $\kappa$-Poincaré quantum group, denoted by $\mathbb{C}_{\kappa}[ISO({\,1,1})]$. This notation characterizes $\kappa$-Poincar\'e as a noncommutative deformation of the Hopf algebra $\mathbb{C}_{\kappa}[ISO({\,1,1})]$ of complex functions on the Poincar\'e group $ISO({\,1,1})$. The algebra sector reads~\cite{Mercati:2023soon} (all greek indices run in the set $\{+,-\}$)
   \begin{equation} \label{eq:galgebra}
    \begin{aligned}
    &[\Lambda^{\mu}_{\!\ \!\ \nu},\Lambda^{\rho}_{\!\ \!\ \sigma}]=0, \quad [a^{\mu},a^{\nu}]=i(v^{\mu}a^{\nu}-v^{\nu}a^{\mu})\\
    &[a^{\gamma},\Lambda^{\mu}_{\!\ \!\ \nu}]=i[(\Lambda^{\mu}_{\!\ \!\ \alpha}v^{\alpha}-v^{\mu})\Lambda^{\gamma}_{\!\ \!\ \nu}+(\Lambda^{\alpha}_{\!\ \!\ \nu}g_{\alpha \beta}-g_{\nu \beta})v^{\beta}g^{\mu\gamma}]\\
    &\Lambda^{\mu}_{\!\ \!\ \alpha}\Lambda^{\nu}_{\!\ \!\ \beta}g^{\alpha \beta}=g^{\mu\nu}, \quad \Lambda^{\rho}_{\!\ \!\ \mu}\Lambda^{\sigma}_{\!\ \!\ \nu}g_{\rho \sigma}=g_{\mu\nu},
    \end{aligned}
\end{equation}
where  
\begin{equation}
\label{eq:algspec}
 v^\mu=(2,0) \,, \qquad    g_{\mu\nu} = \left(
\begin{array}{cc}
    0 & 1 \\
    1 & 0
\end{array}
    \right) \qquad \Rightarrow \qquad  g_{\mu\nu}v^\mu v^\nu=0 \,.
\end{equation}

The coproduct $\Delta$, antipode $S$ and counit $\epsilon$, which codify information on the quantum group product, inversion and identity are un-deformed, and their expressions are given by 
\begin{equation} \label{eq:gcoalgebra}
    \begin{aligned}
    &\Delta[\Lambda^{\mu}_{\!\ \!\ \nu}]=\Lambda^{\mu}_{\!\ \!\ \alpha}\otimes \Lambda^{\alpha}_{\!\ \!\ \nu}, \quad \Delta[a_{\mu}]=\Lambda^{\mu}_{\!\ \!\ \nu} \otimes a^{\nu}+a^{\mu}\otimes 1 \\
    &S[\Lambda^{\mu}_{\!\ \!\ \nu}]=(\Lambda^{-1})^{\mu}_{\!\ \!\ \nu}, \quad S[a^{\mu}]=-(\Lambda^{-1})^{\mu}_{\!\ \!\ \nu}a^{\nu}, \quad \epsilon[{\Lambda^{\mu}_{\!\ \!\ \nu}}]=\delta^{\mu}_{\nu}, \quad \epsilon[a^{\mu}]=0.
    \end{aligned}
\end{equation}

The Poincar\'e transformations of spacetime coordinates can be understood in terms of a left co-action operator $\cdot \, {}' :\mathcal{A}\rightarrow \mathbb{C}_{\kappa}[ISO({\,1,1})] \otimes \mathcal{A}$. We will write this co-action in the following compact way:
\begin{equation} \label{eq:coaction}
    x'{}^{\mu}=\Lambda^{\mu}_{\!\ \!\ \nu}x^{\nu}+a^{\mu} \, ,
\end{equation}
where the product on the right-hand side is understood as the tensor product $\Lambda^{\mu}_{\!\ \!\ \nu} \otimes x^{\nu}+a^{\mu} \otimes 1$. In this notation, it is understood that $[\Lambda^{\mu}_{\!\ \!\ \nu},x^{\rho}]=[a^{\mu},x^{\nu}]=0$. It is easy to check that, given the coordinate transformation~\eqref{eq:coaction} and  the commutation rules~\eqref{eq:galgebra}, the commutator \eqref{eq:cnc} is left invariant, in the sense that
\begin{equation}
    [x'^+,x'^-]=2i \, x'^- \,,
\end{equation}
and the commutation relations appear identical to all inertial observers. 
The symmetries can also be described in terms of the dual Hopf Algebra $\mathcal{U}_\kappa[\mathfrak{iso}(1,1)]$, which can be thought of as a non-commutative deformation of the universal enveloping algebra $\mathcal{U}[\mathfrak{iso}(1,1)]$ of the Poincaré Lie algebra $\mathfrak{iso}(1,1)$. To extract the relevant structures of $\mathcal{U}_\kappa[\mathfrak{iso}(1,1)]$, we apply a finite transformation on non-commutative plane waves, with a given ordering, and extract the action of the generators of the algebra by evaluating the first order of the transformation rules of plane waves. In this calculation and throughout the manuscript, we choose to work with the $x^+$ to-the-right ordering, and a transformed plane wave can be written as
\begin{equation}
    \label{eq:tpw}
    e^{ik_-{x'}^-}e^{ik_+{x'}^+} \, ,
\end{equation}
where ${x'}^-,{x'}^+$ can be read off from \eqref{eq:coaction} and $k_\mu\in\mathbb{C}$.\footnote{Here and in the following, we consider ordered exponentials of the non-commutative coordinates with both real and complex parameters. The properties of the exponentials do not depend whether the parameters are real of complex, in general. For the sake of simplicity, we will refer to these functions as \textit{plane waves}, even if the parameter has an imaginary component (see \cref{sec:oldandnew})}.
From the last line of \eqref{eq:galgebra}, specified by \eqref{eq:algspec}, the Lorentz part of the transformation can be parametrized by a single operator $\tau$, as follows: 
\begin{equation}
    \Lambda^\mu{}_\nu = \left( \begin{array}{cc}
         e^{\tau} &  0 \\
         0 & e^{-\tau} 
    \end{array} \right) \,.
\end{equation}
From commutators \eqref{eq:galgebra}, it is possible to show that
\begin{equation}
\label{eq:paramcomm}
    [a^+,\tau]=2i(e^\tau-1) \,, \qquad [a^-,\tau]=0 \, . 
\end{equation} 
Using relations \eqref{eq:paramcomm} and techniques developed in \cite{Lizzi:2021rlb}, we can write \eqref{eq:tpw} as 
\begin{equation}
    e^{ik_-e^{-\tau}x^-}e^{\frac{i}{2}\log[1+e^\tau(e^{2k_+}-1)]x^+}e^{ik_-a^-}e^{ik_+a^+} \, . 
\end{equation}
Focusing on the Lorentz sector of the transformation, we want to write 
\begin{equation}
\label{eq:boostexp}
    e^{ik_-e^{-\tau}x^-}e^{\frac{i}{2}\log[1+e^\tau(e^{2k_+}-1)]x^+}\approx \qty(1+i\tau N\triangleright) e^{ik_-x^-}e^{ik_+x^+}+\mathcal{O}(\tau^2) \, ,
\end{equation}
where $N$ is the boost operator in $\mathcal{U}_\kappa[\mathfrak{iso}(1,1)]$ and $\triangleright$ is a left action $\triangleright : \mathcal{U}_\kappa[\mathfrak{iso}(1,1)] \otimes \mathcal A  \to \mathcal A$. Expanding the left hand-side at first order in $\tau$, one finds
\begin{equation}
    e^{ik_-e^{-\tau}x^-}e^{\frac{i}{2}\log[1+e^\tau(e^{2k_+}-1)]x^+}\approx \qty(1-i\tau x^-k_-)e^{ik_-x^-}\qty(1+i\tau x^+\qty(\frac{1-e^{-2k_+}}{2}))e^{ik_+x^+}\, .
\end{equation}
This can be understood as a non-linear deformation of the action of the standard boost operator on a commutative plane-wave which is then mapped to a non-commutative one with a given ordering (in this case $x^+$ to the right), by means of a Weyl map $\Omega: C[\mathbb{R}^2] \rightarrow\mathcal{A}$~\cite{Agostini:2003vg}, defined as ($kx$ is a shorthand for $k_\mu x^\mu$):
\begin{equation}
\label{eq:ncpws}
    \Omega(e^{ikx})=e^{ik_-x^-}e^{ik_+x^+} \, .
\end{equation}
The action of the boost operator $N$ can thus be written as 
\begin{equation}
\label{eq:nlboost}
    N\triangleright \Omega(e^{ikx})=\Omega\qty[\qty((ix^-\partial_-)+x^+\qty(\frac{1-e^{2i\partial_+}}{2}))e^{ikx}] \, .
\end{equation}
By inspecting the translation sector of the transformation, the action of the $\Tilde{P_{\pm}}$ generators can be defined as 
\begin{equation}
\label{eq:translpw}
    \Tilde{P}_{\pm}\triangleright \Omega(e^{ikx}) = \Omega(-i\partial_\pm e^{ikx})=k_\pm\Omega( e^{ikx})
\end{equation}
Using expressions \eqref{eq:nlboost}-\eqref{eq:translpw} and applying the generators in succession on a single plane waves, one finds the commutators: 

\begin{equation}
\label{eq:aalgebra}
    [N,\Tilde{P}_+]=i\qty(\frac{1-e^{-2\Tilde{P}_+}}{2}) \qquad [N,\Tilde{P}_-]=-i\Tilde{P}_- \, , 
\end{equation}
which can be easily shown to satisfy the Jacobi identities.
The coproducts encode the deviation from the Leibniz rule, and are found by applying the generators on products of plane waves. The antipode is obtained by acting on ``inverse'' plane waves, \textit{i.e.} plane waves which multiplied by their standard counterpart give the identity. The counit codifies the action of the generators on plane waves with $k=0$.
\begin{equation} \label{eq:acoalgebra}
    \begin{aligned}
    &\Delta[\Tilde{P}_+]=\Tilde{P}_+\otimes 1+ 1\otimes \Tilde{P}_+, \quad \Delta[\Tilde{P}_-]=\Tilde{P}_- \otimes 1+e^{-2\Tilde{P}_+}\otimes \Tilde{P}_- \\
    &\Delta[N]=N \otimes 1+e^{-2\Tilde{P}_+}\otimes N , \quad S[N]=-Ne^{2\Tilde{P}_+}, \quad S[\Tilde{P}_-]=-\Tilde{P}_-e^{2\Tilde{P}_+}\\
    & S[\Tilde{P}_+]=-\Tilde{P}_+,\quad 
    \epsilon[N]=\epsilon[\Tilde{P}_+]=\epsilon[\Tilde{P}_-]=0 \,.
    \end{aligned}
\end{equation}
The procedures outlined above define a Hopf Algebra: all its axioms are satisfied, including the compatibility rules with the commutators~\eqref{eq:aalgebra} (\textit{i.e.}, the homomorphism property of $\Delta$, $S$ and $\epsilon$). The structures thus obtained
define the lightlike $\kappa$-Poincaré Hopf algebra in the so-called bicrossproduct basis (characterized by momenta which close a Hopf subalgebra~\cite{Majid:1994cy,Ballesteros:1996awf}). In particular, expressions \eqref{eq:nlboost},\eqref{eq:translpw} define the infinite-dimensional representation of $\mathcal{U}_\kappa[\mathfrak{iso}(1,1)]$ in the bicrossproduct basis.

The mass Casimir element of this algebra is 
\begin{equation}
\label{eq:nlcas}
    C=\frac{1}{2}\Tilde{P}_-(e^{2\Tilde{P}_+}-1) \, .
\end{equation}

The action of the Weyl map is also useful to define generic non-commutative functions in $\mathcal{A}$, by means of a non-commutative Fourier transform 
\begin{equation}
    f(x)=\int \dd^2 k \, \Tilde{f}(k) \Omega(e^{ikx}) \, .
\end{equation}
For such generic functions, a $\kappa$-Poincaré transformation can be written as ($id$ is the identity map, and the dots indicate all higher order monomials in the transformation parameters, with a given, specified ordering: in this case, $\tau$ is chosen to be to the right of $a^+$, which is, in turn, to the right of $a^-$):
\begin{equation}
\label{eq:Tmatrix}
    f(x') = e^{i \, a^- \otimes \Tilde{P_-}}   e^{i \, a^+ \otimes \Tilde{P_+}} e^{i \, \tau\otimes \, N}  (id \otimes \triangleright) f(x) = 1 \otimes f(x) + i \, a^\mu \otimes  \Tilde{P}_\mu \triangleright f(x) +  i \, \tau \otimes N  \triangleright f(x) + \dots
\end{equation}
with $\Tilde{P}_\mu,N\in \mathcal{U}_\kappa[\mathfrak{iso}(1,1)]$ and the left action on coordinates is easily read from \eqref{eq:nlboost},\eqref{eq:translpw},
\begin{equation}
\label{eq:leftaction}
    \Tilde{P}_\mu \triangleright x^\nu = - i \, \delta^\mu{}_\nu \,, \qquad N \triangleright x^\pm  = \pm i x^\pm \,. 
\end{equation}

A peculiarity of the $\kappa$-lightlike Hopf algebra, which will prove to be useful in characterizing the physical results of this work is the fact it is quasi-triangular,\footnote{In our specific case, a stronger condition holds: the R-matrix is \textit{triangular,} meaning that $R^{BA}_{DC}R^{CD}_{EF} = \delta^A{}_E \, \delta^B{}_F$~\cite{Ballesteros_1995}.} \textit{i.e.}, it admits a quantum $R$-matrix. It has been derived in \cite{Ballesteros_1995,BALLESTEROS_1996,Ballesteros:1996awf}, by exploiting an isomorphism between
$\mathcal{C}_\kappa(ISO(d,1))$ and $\mathcal{U}_\kappa(\mathfrak{iso}(d,1))$, where $d=1,2,3$. In $1+1$ dimensions, the expression of the R-matrix is given by 
\begin{equation}
\label{eq:Rmat}
    R=e^{-2i\Tilde{P}_+\otimes N}e^{2i N\otimes \Tilde{P}_+} \, , 
\end{equation}
and in terms of it, relations \eqref{eq:galgebra}, specified by  \eqref{eq:algspec} can be written in a compact way as ``RTT" relations, often used in the quantum group literature \cite{wess1999qdeformed}. Relations \eqref{eq:leftaction} define a three-dimensional representation $\rho^A_B$, with  $A,B=\{+,-,2\}$ for $\Tilde{P}_\mu,N$ acting on vectors of the form $X^A\equiv (x^\mu,1)$:
  \begin{equation}
\rho(\Tilde{P}_+)^A{}_B = \left( \begin{array}{ccc}
0 & 0  & -i  \\
 0  &  0 & 0 \\
 0  &  0 & 0 
\end{array}\right) \,,
    \qquad
\rho(\Tilde{P}_-)^A{}_B =  \left( \begin{array}{ccc}
0 & 0  & 0  \\
 0  &  0 & -i \\
 0  &  0 & 0 
\end{array}\right)\,,
\qquad
\rho(N)^A{}_B = \left( \begin{array}{ccc}
-i & 0  & 0  \\
 0  &  +i & 0\\
 0  &  0 & 0 
\end{array}\right)\,.
    \end{equation}
By noticing that the $\rho(\Tilde{P}_\pm)^A_B$ matrices are nilpotent, \eqref{eq:Rmat} reduces to:
\begin{equation}
   R=1\otimes 1 -2i\Tilde{P}_+\otimes N + 2i N\otimes \Tilde{P}_+\, ,
\end{equation}
and by realizing the tensor product as the standard Kronecker product, in components we find
\begin{equation}
\label{eq:Rmatcomp}
\begin{aligned}
        R^{AB}_{CD}&= \delta^{A}_{C}\delta^{B}_{D}+2 \, i \, \left( \delta^{A}_+\delta^{B}_+\delta^{2}_C\delta^{+}_D - \delta^{A}_+\delta^{B}_-\delta^{2}_C\delta^{-}_D - \delta^{A}_+\delta^{B}_+\delta^{+}_C\delta^{2}_D + \delta^{A}_-\delta^{B}_+\delta^{-}_C\delta^{2}_D \right)
\end{aligned}
 \, .
\end{equation}
By defining 
\begin{equation}
    T^A{}_B = \left( \begin{array}{ccc}
\Lambda^+{}_+ & \Lambda^+{}_-  & a^+  \\
 \Lambda^-{}_+  &  \Lambda^-{}_- & a^- \\
 0  &  0 & 1 
\end{array}\right) = \left( \begin{array}{ccc}
e^\tau  & 0  & a^+  \\
0  &  e^{-\tau} & a^- \\
 0  &  0 & 1 
\end{array}\right) \,,
\end{equation}
one can explicitly verify that the expression 
\begin{equation}
\label{eq:RTT}T^A{}_C \, T^B{}_D \, R^{DC}_{EF} = R^{BA}_{CD} \, T^C{}_E \, T^D{}_F \,,
\end{equation}
reproduces the commutation relations \eqref{eq:galgebra}. Moreover, the commutation relations between coordinates can be written in a compact way as
\begin{equation}
\label{eq:RXX}
    X^A \,  X^B = R^{BA}_{CD} \, X^C \, X^D \, . 
\end{equation}
Equivalently, these ``RXX" relations can also be verified using the infinite-dimensional representation of $\mathcal{U}_\kappa(\mathfrak{iso}(d,1))$ and relations \eqref{eq:leftaction}. For instance: 
\begin{equation}
    x^+x^-=\mu\circ R\triangleright (x^-\otimes x^+)=x^-x^++2i \,, 
    \end{equation}
where $\mu:\mathcal A \otimes \mathcal A \to \mathcal A$ is the noncommutative multiplication of $\mathcal A$. This deformed flip property can also be extended to products of plane waves. Indeed, one can verify that
\begin{equation}
\label{eq:pwswitch}
    \mu \circ R \triangleright  \left[   \Omega( e^{i q x})  \otimes  \Omega( e^{i k x}) \right] = \Omega( e^{i k x}) \Omega( e^{i q x})\, , 
\end{equation}
where $k,q\in \mathbb{C}^2$. As we will see in the subsequent sections, the R-matrix proves to be a valuable instrument in switching plane waves even in the braided tensor product algebra introduced in \cite{Lizzi:2021rlb,DiLuca:2022idu}, so that the task of covariant quantization of the non-commutative scalar field becomes more feasible. 

In what follows, we will work in a different basis for $\mathcal{U}_\kappa[\mathfrak{iso}(1,1)]$, which is connected to the bicrossproduct one by a redefinition of the $+$ momentum, given by 
\begin{equation}
\label{eq:linmom}
    P_+=\frac{1}{2}\left(e^{2 \Tilde{P}_+}-1\right) \, . 
\end{equation}
The $\mathcal{U}_\kappa[\mathfrak{iso}(1,1)]$ commutators are, in this basis, the undeformed  ones of the Poincaré algebra,
\begin{equation}
\label{eq:lincomm}
    [N,P_+]=iP_+ \qquad [N,P_-]=-iP_- \,.
\end{equation}
All the non-linearity is moved to the coproducts and the antipodes, which now take the form 
\begin{equation}
    \begin{aligned}
    \label{eq:lincop}
        &\Delta[P_+]=P_+\otimes 1 + 1\otimes P_+ + 2 P_+\otimes P_+, \qquad S(P_+)=-\frac{P_+}{1+2P_+},\\
        &\Delta [P_-]=P_-\otimes 1 + 1 \otimes P_- - \frac{2P_+}{1+2P_+}\otimes P_-, \qquad S(P_-)=-P_-(1+2P_+),\\
        &\Delta[N]=N\otimes 1 + 1 \otimes N - \frac{2P_+}{1+2P_+}\otimes N, \qquad S(N)=-N(1+2P_+) \, , 
    \end{aligned}
\end{equation}
while the counits remain all zero. The Casimir element in these variables can be obtained by substituting \eqref{eq:linmom} in \eqref{eq:nlcas} and is undeformed:
\begin{equation}
\label{eq:lincas}
    C=P_+P_- \, , 
\end{equation}
in agreement with the linearity of commutators \eqref{eq:lincomm}.

Nonlinear transformations of the translation generators lead to different bases for the $\mathcal{U}_\kappa[\mathfrak{iso}(1,1)]$ Hopf algebra, which, as we will see in the following, correspond to different coordinate systems on momentum space.\footnote{Notice that Hopf algebras are structures that are invariant under general nonlinear changes of basis \cite{majid_2002}. This is in stark contrast with Lie algebras, which can only be meaningfully said to be invariant under linear transformations of the basis. This enlarged invariance is guaranteed by the additional structures like coproduct, counit and antipode. One can usually change the commutators of a Hopf algebra to make them take almost any desired form, but only at the cost of changing the coproducts, etc.} The theory we are presenting in this paper has the aspiration of being invariant under general coordinate transformations on momentum space. This would imply that the physical observables do not depend on the momentum coordinate systems used in their prediction (see the discussion in \cite{Amelino-Camelia:2011lvm}, Sec.~II). The presence of such an invariance in our theory is supported by the preliminary results in \cite{DiLuca:2022idu} (sec. 2.3), which show that the two-point functions of the theory  are the same regardless of the coordinate system on momentum space that was used to calculate them.

Even in a generally-covariant theory, certain situations are better described by certain choices of coordinates, \textit{e.g.} Cartesian coordinates in Minkowski space are preferred because they transform covariantly under Lorentz transformations. In our model, the choice of coordinates $P_{\pm}$ has the same advantage: they transform in an undeformed fashion under boosts. For this reason, we find it convenient to work with them. As it turns out, plane waves are eigenfunctions of the momenta, and the $P_\pm$ basis has the following eigenvalues:
\begin{equation}
    P_+\Omega(e^{ikx})=\frac{1}{2}(e^{2k_+}-1)\Omega(e^{ikx}) \qquad P_-\Omega(e^{ikx})=k_-\Omega(e^{ikx}) \,.
\end{equation}
These relations inspire a redefinition of the momenta appearing in the plane waves:
\begin{equation}
\label{eq:linmomdef}
    \xi_-=k_- \,, \qquad \xi_+=\frac{1}{2}(e^{2k_+}-1) \,,  \qquad \Rightarrow\qquad  \Omega (e^{i \, k x}) = e^{i\xi_-x^-}e^{\frac{i}{2}\ln(1+2\xi_+)x^+} \,,
\end{equation}
so that
\begin{equation}
    \label{eq:Edefinition}P_{\pm}\Omega(e^{ikx})=P_{\pm}\triangleright e^{i\xi_-x^-}e^{\frac{i}{2}\ln(1+2\xi_+)x^+}=\xi_{\pm}E[\xi] \, ,
\end{equation}
where we defined $E[\xi]\equiv e^{i\xi_-x^-}e^{\frac{i}{2}\ln(1+2\xi_+)x^+}$. 
The algebraic properties of non-commutative plane waves reflect the non-linear structure of momentum space.
First of all, from redefinition \eqref{eq:linmomdef}, we notice that the $\xi_+$ component of the momentum is bounded from below, so that plane waves $E[\xi]$ only cover half of momentum space \cite{Lizzi:2021rlb,DiLuca:2022idu}.
Products of plane waves define the deformed composition law for momentum (denoted by $\Delta)$ and the deformed inverse momenta (denoted with $S$), which mimick the structures of coproduct and antipode, respectively: 
\begin{equation}
\label{eq:linoplus}
    \begin{aligned}
&\Delta(\xi ,\eta)=\left(\xi_- +\frac{\eta_-}{1+2\xi_+},\xi_++\eta_++2\xi_+\eta_+\right)\\
&S(\xi) = \left(-\xi_-(1+2\xi_+),-\frac{\xi_+}{1+2\xi_+}\right)
     \end{aligned}      
     \, , 
\end{equation}
for $\xi,\eta\in\mathbb{C}^2$. The Hopf algebra properties then imply the following consistency relations between composition law and antipode, which can be checked explicitly using \eqref{eq:linoplus}: 
\begin{equation}
\label{eq:haprop}
\begin{aligned}
    &\Delta[\xi, \Delta (\eta , \chi) ] = \Delta[\Delta(\xi,\eta) , \chi] = \Delta[\xi , \eta , \chi] \,, 
    \\
    &\Delta[\xi, S(\xi)] =  \Delta[S(\xi), \xi] = 0 \,,
    \\
    &S[\Delta(\xi,\eta)] = \Delta[S(\eta), S(\xi)] \,,
\end{aligned}
\end{equation}
for $\xi,\eta,\chi\in\mathbb{C}^2$. The first relation implies the associativity of the composition law, the second the existence of a momentum inverse, and the third codifies the anti-homomorphism property of the antipode. The Casimir element \eqref{eq:lincas} defines mass-shells in momentum-space through the constraint
\begin{equation}
\label{eq:stdos}
    m^2=\xi_+\xi_- \,  ,
\end{equation}
just as in the ordinary theory. 

Later on, we will see that the set of operators $P_+,P_-,N$ can also be represented in terms of creation-annihilation operators of standard quantum field theory, and their action on well defined (multi)-particle states follows the Hopf algebraic structures displayed above. Before diving into such considerations, we review some of the basics elements needed to construct a consistent quantum field theory on the $1+1$D lightlike $\kappa$-Minkowski quantum space-time, following \cite{Lizzi:2021rlb,DiLuca:2022idu}.  

\subsection{Braided N-point algebra and its representations}

The construction of the $1+1$D braided lightlike $\kappa$-Minkowski algebra $\mathcal{A}^{\tilde \otimes N}$ was developed in \cite{Lizzi:2021rlb}. The defining commutation relations are given by
\begin{equation} \label{eq:1+1alg1}
    [x^+_a,x^+_b]=2i(x^+_a-x^+_b), \quad [x^+_a,x^-_b]=2ix^-_b, \quad [x^-_a,x^-_b]=0 \, ,
\end{equation}
with $a,b=1,...,N$. These relations can equivalently be written in terms of center of mass and relative coordinates:
\begin{equation}
\label{eq:comrel}
   x^{\mu}_{cm}=\frac{1}{N}\sum_{a=1}^N x^{\mu}_{a} \qquad y^{\mu}_{a}=x^{\mu}_{a}-x^{\mu}_{cm} \,  , 
\end{equation}
so that \eqref{eq:1+1alg1} becomes 
\begin{equation} \label{eq:1+1alg2}
    [x^+_{cm},x^-_{cm}]=2ix^-_{cm}, \quad [x^+_{cm},y_a^{\pm}]=\mp 2i y_a^{\pm}, \quad [x^-_{cm},y_a^{\pm}]=0.
\end{equation}
It is easy to check that the coordinate differences $\Delta x^\mu_{ab}\coloneqq x^\mu_a-x^\nu_b$ are commutative (a feature also shared by the braided tensor product of the $\theta$-Moyal non-commutative space-time \cite{Fiore:2007vg}): 
\begin{equation}
    [\Delta x^{\mu}_{ab},\Delta x^{\nu}_{cd}]=0, \quad a,b,c,d=1,\dots, N \,\,\,\text{ and } \mu,\nu=+,-.
\end{equation}
This, combined with the fact that $\kappa$-Poincaré invariant $N$-point functions depend solely on coordinate differences (proved in \cite{Lizzi:2021rlb}), immediately tells us that $\kappa$-Poincaré invariant $N$-point functions are commutative themselves. This greatly simplifies the interpretation of the theory, given that all physical information should be encoded in $N$-point functions.
Once again, commutation relations \eqref{eq:1+1alg1} can be written in terms of an $R$-matrix \cite{wess1999qdeformed} as 
\begin{equation}
   X^A_a \,  X^B_b = R^{BA}_{CD} \, X^C_b \, X^D_a \,, 
\end{equation}
where $X^A_a=(x^{\mu}_a,1)$, and the operators appearing in the $R$-matrix act in the same way on the $x^\mu_a$ coordinates whatever the value of $a$.

In \cite{Lizzi:2021rlb}, a representation for the center of mass and relative coordinates has been found, and reads
\begin{equation}
\label{eq:hermrep}\begin{aligned}
    &\hat{x}^+_{cm}=2ix^-_{cm}\frac{\partial}{\partial {x^-_{cm}}}+i+2i\sum_{a=1}^{N-1}\bigg(y_a^+\frac{\partial}{\partial y_a^+}-y_a^-\frac{\partial}{\partial y_a^-}\bigg), \quad
  &\hat{x}^-_{cm}=x^-_{cm}, \quad \hat{y}_a^+=y_a^+, \quad \hat{y}_a^-=y_a^-, 
\end{aligned}\end{equation}
and is such that $x_{cm}^{\pm},y_a^{\pm}$ are Hermitian. 
For purposes which shall be clear once we discuss non-commutative plane waves in more detail, we will consider a more general, one-parameter class of representations, given by 
\begin{equation}\label{eq:paramrep}
\begin{aligned}
    &\hat{x}^+_{cm}= 
    2i \left( x^-_{cm}\frac{\partial}{\partial {x^-_{cm}}} + s    \right) +2i\sum_{a=1}^{N-1}\bigg(y_a^+\frac{\partial}{\partial y_a^+}-y_a^-\frac{\partial}{\partial y_a^-}\bigg), \quad
  &\hat{x}^-_{cm}=x^-_{cm}, \quad \hat{y}_a^+=y_a^+, \quad \hat{y}_a^-=y_a^-, 
\end{aligned}
\end{equation}
where $0<s<1$, and the Hermitian representation is regained with $s= 1/2$. When analyzing plane waves, in the subsequent sections, it is useful to study the action of operators of the type $e^{i t x_{cm}^+}$ on functions of the braided tensor product algebra. Using \eqref{eq:paramrep}, it is easy to check that, for any complex $t$, 
\begin{equation} \label{eq:action+}
  e^{i t \hat{x}^+_{cm}}f(x^-_{cm},y_a^+,y_a^-)=e^{-t \, a}f(e^{-2 t}x^-_{cm},e^{2 t}y_a^+,e^{-2 t}y_a^-)  \,.
\end{equation}
Then, exploiting the fact that 
\begin{equation}
    e^{i t x_a^+}=e^{i t (x^+_{cm}+y_a^+)}=e^{i \left(\frac{e^{2 t}-1}{2} \right)y_a^+}e^{i t x^+_{cm}} \,,
\end{equation}
we obtain 
 \begin{equation}
e^{- \frac{\pi}{2} {x}^+_a} =e^{i \left(\frac{e^{\pi \, i}-1}{2} \right)y_a^+}e^{-\frac{\pi}{2}x^+_{cm}} =e^{- i \, y_a^+} e^{-\frac{\pi}{2} {x}^+_{cm}} \,,
 \end{equation}
so that
\begin{equation}
\label{eq:operatorrefl}
    e^{-\frac{\pi}{2}\hat{x}^+_a}f(x^-_{cm},y_a^+,y_a^-)=e^{-iy_a^+}e^{-i \pi\, s}f(e^{-i\pi }x^-_{cm},e^{i\pi }y_a^+,e^{-i\pi }y_a^-)=e^{-i \pi s} e^{-iy_a^+}f(-x^-_{cm},-y_a^+,-y_a^-) \,.
\end{equation}
The square of this operator is then simply 
\begin{equation}
    e^{- \pi \, \hat{x}^+_a}f(x^-_{cm},y_a^+,y_a^-) = e^{-2 i \pi \, s} f(x^-_{cm},y_a^+,y_a^-) \,.
\end{equation}
and thus
 \begin{equation} \label{eq:repprx+}
 e^{-n\pi {x}^+_a}f(x^-_{cm},y_a^+,y_a^-) = e^{- 2 i \pi \, s \,n} f(x^-_{cm},y_a^+,y_a^-) \,.
 \end{equation}

Having introduced the one-parameter family of representations~\eqref{eq:paramrep}, we would like to find a condition that fixes the parameter $s$. This will be identified in the next Section, in order to eliminate a sign ambiguity that emerges when introducing a certain type of noncommutative plane waves (first introduced in \cite{Bevilacqua:2022fbz,DiLuca:2022idu}) that are necessary to ensure the covariance of the theory. In the meantime, we need to briefly discuss the Hermiticity/self-adjointness properties of the N-point coordinate operators $x_a^\mu$. This will be necessary, as later we will need to introduce an involution that sends a noncommutative plane wave into its inverse, which is necessary in order to discuss field theory.
What we would like is an involutive, anti-linear anti-homomorphism which sends $E[\xi]$ 
(from Eq.~\eqref{eq:Edefinition}) into $E^\dagger[\xi]$ such that $E^\dagger[\xi] E[\xi] =E[\xi] E^\dagger[\xi]=1 $. We start by defining a putative operator $*$ as the "naive" Hermitian conjugation on operators, such that its action on $\hat{x}^+_{cm}$ is given by  
\begin{equation}
(\hat{x}_{cm}^{+})^{*}=\hat{x}_{cm}^++2i(2s-1)
\end{equation}
where, as expected, we obtain that $(\hat{x}_{cm}^+)^* = \hat{x}_{cm}^+$  only when $s=1/2$, which corresponds to the symmetric ordering for the representation~\eqref{eq:paramrep}. We can now define $\dagger$ as the operator that leaves $\hat{x}_{cm}^+$ invariant for any choice of $s$, $(\hat{x}_{cm}^+)^\dagger=\hat{x}_{cm}^+$, so that its relation with $*$ is simply given by
\begin{equation}\label{eq:NewHermitianConjugate}
    (\hat{x}_{cm}^+)^\dagger=(\hat{x}_{cm}^+)^*-2i(2s-1)
\end{equation}
The $(\cdot)^*$ operator is the Hermitian conjugate with respect to the standard inner product of $L^2(\mathbbm{R}^{2N-1})$:
\begin{equation}
 \int_{\mathbbm{R}^{2N-1}} \bar{\psi} ~ \varphi ~  dx_{cm}^- dy_1^-\dots dy_{N-1}^- dy_1^+\dots dy_{N-1}^+ \,,
\end{equation}
where $\psi,\phi$ are square-integrable functions on $\mathbbm{R}^{2N-1}$. The $\dagger$ operation is the Hermitian conjugate with respect to a different inner product:
\begin{equation}
\label{eq:misurabuona}
  \int_{\mathbbm{R}^{2N-1}} (x_{cm}^-)^{2s-1} \, \bar{\psi} \, \varphi \,  dx_{cm}^- dy_1^-\dots dy_{N-1}^- dy_1^+\dots dy_{N-1}^+ \,,
\end{equation}
where, in this case, the space of functions that have a finite norm is different from $L^2(\mathbbm{R}^{2N-1})$. For $s > 1/2$, it includes $L^2(\mathbbm{R}^{2N-1})$, and also functions that diverge sufficiently slowly in $x_{cm}^- \to 0$. For $s < 1/2$, the space is smaller than $L^2(\mathbbm{R}^{2N-1})$, as the functions need to go to zero sufficiently fast at $x_{cm}^- \to 0$. The fact that the representations of $\mathcal{A}$ and the related braided algebras require different inner products for the self-adjointness of the generators has been already noticed in~\cite{Dabrowski:2010yk,Lizzi:2018qaf,Lizzi:2019wto}. 
From now on, we will use the $\dagger$ operator to conjugate plane waves.

\section{Braided lightlike $\kappa$-deformed QFT}

\subsection{Old and new-type noncommutative plane waves and momentum space}\label{sec:oldandnew}

We have introduced plane waves for a single copy of the $\mathcal{A}$ in \cref{sec:kspaceandgroup} using the linear momentum parametrization. As it will be relevant for what follows, we add a label indicating in which copy of the braided tensor product algebra $\mathcal{A}^{\tilde \otimes N}$ the plane wave is defined: 
\begin{equation}
\label{eq:otpw}
    E_a[\xi]=e^{i\xi_-x_a^-}e^{\frac{i}{2}\ln(1+2\xi_+)x_a^+} \, . 
\end{equation}
Under the involution that leaves $x^+$ invariant, the above transformes as
\begin{equation}
\label{eq:otdagger}
    E^\dagger_a[\xi]=E_a[S(\xi)]  \, , 
\end{equation}
where $S(\xi)$ is the antipode defined in \eqref{eq:linoplus}, while the product of two plane waves gives a composition law compatible with the coproducts in \eqref{eq:lincop}:
\begin{equation}
    E_a[\xi]E_a[\eta]=E_a[\Delta(\xi,\eta)] \, .
\end{equation}
The space-time coordinates $x^\pm$ close the Lie algebra~\eqref{eq:cnc} of the affine group of the line, $\text{aff}(1)$. Plane waves \eqref{eq:otpw} span the connected component of the identity of the corresponding Lie group, $\text{Aff}(1)$. This is just a semiplane of 1+1-dimensional Minkowski space, bounded by a straigth line~\cite{Lizzi:2021rlb,DiLuca:2022idu}. The boundary is given by the reality constraint for the logarithm term, $\xi_+>-1/2$. This implies that such plane waves only cover half of the Minkowski momentum space, an issue already pointed out in \cite{Lizzi:2021rlb}. There, it was shown that a field theory built from plane waves \eqref{eq:otpw} spoils $\kappa$-Poincaré invariance. This can easily be seen by considering a $\kappa$-Poincaré transformation of \eqref{eq:otpw}:
\begin{equation}
\label{eq:otpwtrans}
E_a'[\xi] = e^{ie^{-\tau}x^-}e^{\frac{i}{2}\ln(1+2e^\tau \xi_+)x^+} \, e^{i\xi_-a^-}e^{\frac{i}{2}\ln(1+2\xi_+)a^+} \, .
\end{equation}
Notice that in this linear parametrization, the boost simply acts as a dilation on $\xi_-,\xi_+$, given the linear structure of the commutators \eqref{eq:lincomm}.
For any value of $\tau$, a positive value of $\xi_+$ remains positive, and we obtain a different group element connected to the identity. When $\xi_+$ is negative, an excessively large boost may result in $e^\tau \xi_+<-1/2$, so that the argument of the logarithm in \eqref{eq:otpw} becomes negative and we obtain a group element not connected to the identity, which we can think of as a plane wave of the form \eqref{eq:otpw} with a complex argument.
In \cite{DiLuca:2022idu}, these ``new type'' plane waves were identified as the missing piece of the puzzle needed to construct a consistent $\kappa$-Poincaré invariant field theory.

Suppose we boost a plane wave of the form \eqref{eq:otpw}, such that $e^\tau\xi_+<-1/2$; then, the logarithm term can be written as 
\begin{equation}
    \ln[-|1+2e^{\tau}\xi_+|]
    = i \, \pi + \ln |1+2e^{\tau}\xi_+|  + 2 \, n \, \pi \, i \,.
\end{equation}
Focusing on the Lorentz transformation sector of \eqref{eq:otpwtrans}: 
\begin{equation}
\label{eq:odpwboost}
    e^{i\, e^{-\tau} \xi_-x^-_a} e^{\frac{i}{2} \ln[-|1+2e^{\tau}\xi_+|] \, x^+_a} 
    = e^{i\, e^{-\tau} \xi_-x^-_a} e^{\frac{i}{2} \ln|1+2e^{\tau}\xi_+| \, x^+_a} e^{ -\frac{\pi}{2} \, x^+_a } e^{- n \, \pi \,x^+_a}.
\end{equation}
We now come to an issue not discussed in  \cite{DiLuca:2022idu}. There, using representation \eqref{eq:hermrep}, a sign ambiguity emerges in \eqref{eq:odpwboost}, due to the fact that $e^{-n\pi x_a^+}\equiv (-1)^n$. In our novel parametric representation \eqref{eq:paramrep}, using the identification $e^{-n\pi x_a^+}\equiv e^{-2i\pi s n}$ from \eqref{eq:repprx+}, \eqref{eq:odpwboost} becomes
\begin{equation}
    e^{i\, e^{-\tau} \xi_-x^-_a} e^{\frac{i}{2} \ln[-|1+2e^{\tau}\xi_+|] \, x^+_a} 
    = e^{i\, e^{-\tau} \xi_-x^-_a} e^{\frac{i}{2} \ln|1+2e^{\tau}\xi_+| \, x^+_a} e^{ -\frac{\pi}{2} \, x^+_a } e^{- 2i\,  \pi s \, n}.
\end{equation}
To avoid the aforementioned sign ambiguity, we may choose $s=1$. Notice that this implies that the coordinates are only  Hermitian with respect to the inner product~\eqref{eq:misurabuona}. Nevertheless, the physical quantities characterizing our quantum field theory (two point functions) are not affected by this choice. From now on, whenever we refer to the (braided or not) $\kappa$-Minkowski coordinate algebra we mean representation~\eqref{eq:paramrep} with the choice $s=1$, and the Hermitian conjugate operator $\dagger$ defined in Eq.~\eqref{eq:NewHermitianConjugate}.
Having solved the sign ambiguity, we have singled out one new type plane wave, among the infinite possibilities arising from crossing the momentum space boundary with a too large boost: 
\begin{equation}
    E_a[\xi]\rightarrow \mathcal{E}[e^{-\tau}\xi_-, \frac{1}{2}\ln|1+2e^\tau\xi_+|] \, ,
\end{equation}
where, as in \cite{DiLuca:2022idu}, we have defined 
\begin{equation}
\label{eq:newpws}
    \mathcal{E}_a[\xi]\coloneqq E_a[\xi]e^{-\frac{\pi}{2}x_a^+}
\end{equation}
The properties of plane waves of the type \eqref{eq:newpws} have been discussed in detail in \cite{DiLuca:2022idu}, in particular all the rules to multiply these plane waves among each other and with old-type plane waves, which are necessary for the discussion of QFT. For what follows, we recall that the on-shell relation for this other half of momentum space is still given by \eqref{eq:stdos}.

\subsection{Covariant quantization and oscillator algebra}
\label{sec:covquant}

We expand a scalar field $\phi(x_a)$ in terms of old type and new type plane waves, as outlined in \cite{DiLuca:2022idu}.
\begin{equation} \label{eq:soluzionetot}
   \phi(x_a)= \int d^2\xi \, \delta(\xi_+\xi_--m^2)\tilde{\phi_1}(\xi)E_a[\xi]+  \int d^2\eta\,\delta(\eta_+\eta_--m^2)\tilde{\phi_2}(\eta)\mathcal{E}_a[\eta] \, . 
\end{equation}
Enforcing the on-shell constraints and recalling the ranges of validity of expressions for both types of plane waves, the field can be expressed as 
\begin{equation}
\label{eq:fe1}
    \begin{aligned}
        \phi(x_a)=&  -\int_{-1/2}^0 \frac{d\xi_+}{2\xi_+}\tilde{\phi_1}(\xi_+)E_a\left[\frac{m^2}{\xi_+},\frac{1}{2}\ln(1+2\xi_+)\right]+ \int_0^\infty \frac{d\xi_+}{2\xi_+}\tilde{\phi_1}(\xi_+)E_a\left[\frac{m^2}{\xi_+},\frac{1}{2}\ln(1+2\xi_+)\right]
        +\\
        -&\int_{-\infty}^{-1/2}\frac{d\xi_+}{2\xi_+}\tilde{\phi_2}(\xi_+)\mathcal{E}_a\left[\frac{m^2}{\xi_+},\frac{1}{2}\ln\left(1+2\xi_+\right)\right]
    \end{aligned}
\end{equation}
For convenience, let us introduce the shorthand notation
\begin{equation}
\label{eq:ospws}
\begin{aligned}
    &e_a(\xi_+)\coloneqq
    \exp\left[i\frac{m^2}{\xi_+}x_a^{-}\right]\exp\left[\frac{i}{2}\ln(1+2\xi_+)x_a^+\right] \qquad \xi_+>-\frac{1}{2} \\
    &\epsilon_a(\xi_+)\coloneqq
    \exp\left[i\frac{m^2}{\xi_+}x_a^{-}\right]\exp\left[\frac{i}{2}\ln(1+2\xi_+)x_a^+\right] \qquad \xi_+<-\frac{1}{2}
\end{aligned}
\end{equation}
for on-shell plane waves. Given expression \eqref{eq:ospws}, we have $e_a^\dagger(\xi_+)=e_a(S(\xi_+))$ and $\epsilon_a^\dagger(\xi_+)=\epsilon_a(S(\xi_+))$. Hereafter, we will only focus on on-shell plane waves. Therefore, to further simplify the notation, we remove the $+$ subscript from linear momentum and implicitly refer to the $+$ component of momenta unless otherwise stated. 
We now make a key observation regarding the antipode function  $S(\xi)$. We notice that $S$ maps the $\xi>0$ region into $-1/2<\xi<0$ region and vice versa. The region $\xi<-1/2$ is mapped onto itself via the application of $S$. This suggests that, while the integral containing old type plane waves may be customarily expanded in terms of both old type plane waves \emph{and} their Hermitian conjugates, for the integral containing new type plane waves, only one between $\mathcal{E}$ and $\mathcal{E}^\dagger$ is needed, otherwise one would be overcounting Fourier modes. As a result, the field expansion \eqref{eq:fe1} now reads
\begin{equation}
\label{eq:fe2}
\begin{aligned}
\phi(x_a)=&\int_0^\infty\frac{d\xi}{2\xi}\left[\frac{1}{2\xi+1}\Tilde{\phi}_1\left(S(\xi)\right)e_a^\dagger(\xi)+\tilde{\phi_1}(\xi)e_a(\xi)\right]+\\
+&\int_{-\infty}^{-\frac{1}{2}}\frac{d\xi}{2\xi(1+2\xi)}\Tilde{\phi}_2\left(S(\xi)\right)\epsilon_a^\dagger(\xi)
\end{aligned}
\end{equation}
Taking inspiration from the undeformed QFT, we define
\begin{equation}
\label{eq:FTredef}
    \begin{cases}
        \Tilde{\phi}_1(S(\xi))=a(\xi) \qquad \xi>0\\
        \Tilde{\phi}_1(\xi)=\bar{b}(\xi) \qquad \xi>0\\
        \Tilde{\phi}_2(\xi)=\alpha(\xi) \qquad \xi<-\frac{1}{2} \, , 
    \end{cases}
\end{equation}
where the bar indicates complex conjugation. Upon quantization, $a(\xi)$ will play the role of a particle annihilation operator while $b^\dagger(\xi)$ will play the role of an anti-particle construction operator. The newly introduced operator $\alpha(\xi)$ is defined "across" the momentum space border and introduces a relation between operators $a(\xi)$ and $b^\dagger(\xi)$ when the momenta involved are also across the border. 
The expression for our scalar field is thus
\begin{equation}
    \begin{aligned}
    \label{eq:sfexpr}
\phi(x_a)=&\int_0^\infty\frac{d\xi}{2\xi}\left[\frac{1}{2\xi+1}a(\xi)e_a^\dagger(\xi)+\bar{b}(\xi)e_a(\xi)\right]+\\
+&\int_{-\infty}^{-\frac{1}{2}}\frac{d\xi}{2\xi(1+2\xi)}\alpha(\xi)\epsilon_a^\dagger(\xi)
\end{aligned}
\end{equation}
We now promote the Fourier coefficients $a(\xi),b(\xi),\alpha(\xi)$ and their complex conjugates to operators. The latter will be indicated by the $\dagger$ symbol rather than the bar one \footnote{Although we will indicate the Hermitian conjugates of these operators with the usual $\dagger$ symbol, as is also the case with plane waves, it is important to keep in mind that they act on different Hilbert spaces.}. We adopt a covariant quantization approach, using the Pauli-Jordan function $\Delta_{PJ}(x_1-x_2)$, that is found to be equal to the one employed in the commutative case \cite{DiLuca:2022idu} : one can obtain it starting from the one found in \cite{DiLuca:2022idu} and performing a change of variables to linear momentum. It reads: 
\begin{equation} \label{eq:PJ}
    \begin{aligned}
    \Delta_{\text{PJ}}(x_1-x_2)&=-\int_0^{+\infty}\frac{d\xi}{2\xi}e_1(\xi)e_2^{\dagger}(\xi)+\int_0^{+\infty} \frac{d\xi}{2\xi}\frac{1}{2\xi+1}e_1^{\dagger}(\xi)e_2(\xi)+\\
    &-\int_{-\infty}^{-\frac{1}{2}}\frac{d\eta}{2\eta}\epsilon_1(\eta)\epsilon_2^{\dagger}(\eta) \, . 
    \end{aligned}
\end{equation}
We are now equipped with all the ingredients needed to quantize the scalar field \eqref{eq:fe2},  which we now denote as $\hat{\phi}$ and  treat as an element of $\mathcal{A}\otimes\mathcal{O}(\mathcal{H})$, where $\mathcal{O}(\mathcal{H})$ is the set of operators acting on the (anti-)particle Hilbert space.  The expression for $\hat{\phi}$ is
\begin{equation}
    \begin{aligned}
    \label{eq:sfexprquant}
\hat{\phi}(x_a)=&\int_0^\infty\frac{d\xi}{2\xi}\left[\frac{1}{2\xi+1}a(\xi)e_a^\dagger(\xi)+b^\dagger(\xi)e_a(\xi)\right]+\\
+&\int_{-\infty}^{-\frac{1}{2}}\frac{d\xi}{2\xi(1+2\xi)}\alpha(\xi)\epsilon_a^\dagger(\xi) \, ,
\end{aligned}
\end{equation}
and the covariant quantization rules are 

\begin{equation} \label{eq:quantization}
    [\hat{\phi}(x_1),\hat{\phi}^{\dagger}(x_2)]= \Delta_{\text{PJ}}(x_1-x_2), \quad [\hat{\phi}(x_1),\hat{\phi}(x_2)]=[\hat{\phi}^{\dagger}(x_1),\hat{\phi}^{\dagger}(x_2)]=0.
\end{equation} 
Commutators \eqref{eq:quantization} then involve products of creation and annihilation operators as well as products of non-commutative plane waves. In performing these computations, we must choose a specific ordering for the coordinate functions belonging to different copies of $\mathcal{A}$ in the braided tensor product algebra. Following  \cite{Lizzi:2021rlb,DiLuca:2022idu}, we order the non-commutative plane waves products with the $x_2$ variables to the right. The general strategy to perform this calculation is based on the fact that all plane waves, regardless of their type, can be formally written as in \eqref{eq:ospws}, and they can be exchanged by making use of the $R$-matrix. For any real values of $\xi,\eta$, we have: 
\begin{equation}
\label{eq:ospwflip}
    e_2(\eta)e_1(\xi)=\mu\circ R \triangleright e_1(\xi)\otimes e_2(\eta)=e_1(\xi+2\xi\eta) e_2\qty (\frac{\eta}{1+2\xi+4\xi\eta}) \, , 
\end{equation}
where $R$ is defined in \eqref{eq:Rmat}, and in linear momentum variables reads
\begin{equation}
\label{eq:rmatfinal}
    R=e^{-i\ln(1+2P_+)\otimes N}e^{i N\otimes \ln(1+2P_+)}\, . 
\end{equation}
Whether the waves above are of old or new type depends on the specific values of $\xi,\eta$ considered. This leads to a division of the commutation relations between creation and annihilation in various regions of momentum space. The resulting list of commutation relations is still rather involved, but the overall picture is much simpler than the one presented in \cite{DiLuca:2022idu}, thanks to the linear momentum redefinition.
From the first commutator in \eqref{eq:quantization}, we obtain:

$\bullet$ In the region $\xi\in]0;+\infty[$, $\eta\in]0;\frac{1}{4\xi}[$
\begin{equation}\label{bdb1}
    b^\dagger(\xi)b(\eta)-\frac{1}{1-4\xi\eta}b\left(\frac{\eta+2\xi\eta}{1-4\xi\eta}\right)b^\dagger\left(\frac{\xi+2\xi\eta}{1-4\xi\eta}\right)=-2\xi\delta(\xi-\eta)
\end{equation}

$\bullet$ In the region $\xi\in]0;+\infty[$, $\eta\in]\frac{1}{4\xi};\infty[$
\begin{equation}\label{aad2}
    b^\dagger(\xi)b(\eta)+\frac{1}{1-4\xi\eta}\alpha^\dagger\left(-\frac{\eta+2\xi\eta}{1+2\eta}\right)\alpha\left(-\frac{\xi+2\xi\eta}{1+2\xi}\right)=-2\xi\delta(\xi-\eta)
\end{equation}

$\bullet$ In the region $\eta\in]0;+\infty[$, $\xi\in]0;\infty[$
\begin{equation}
    b^\dagger(\xi)a^\dagger(\eta)=\frac{1+2\eta}{1+2\eta+2\xi}a^\dagger\qty(\eta+2\xi\eta)b^\dagger\qty(\frac{\xi}{1+2\xi+4\xi\eta})
\end{equation}
\begin{equation}
   a(\xi)b(\eta)=\frac{1+2\xi}{1+2\xi+4\xi\eta}b\left(\frac{\eta}{1+2\xi+4\xi\eta}\right)a\left(\xi+2\xi\eta \right)
\end{equation}
\begin{equation}
    a(\xi)a^\dagger(\eta)-\frac{(1+2\xi)(1+2\eta)}{1+2\xi+2\eta}a^\dagger\left(\frac{\eta}{1+2\xi}\right)a\left(\frac{\xi}{1+2\eta}\right)=2\eta(1+2\eta)\delta(\xi-\eta)
\end{equation}

$\bullet$ In the region $\eta\in]-\infty;-\frac{1}{2}[$, $\xi\in]-\infty;-\frac{1}{2}[$
\begin{equation}
    \alpha(\xi)\alpha^\dagger(\eta)+\frac{(1+2\xi)(1+2\eta)}{1+2\xi+2\eta}a^\dagger\left(\frac{\eta}{1+2\xi}\right)a\left(\frac{\xi}{1+2\eta}\right)=2\eta(1+2\eta)\delta(\xi-\eta)
\end{equation}

$\bullet$ In the region $\eta\in]-\infty;-\frac{1}{2}[$, $\xi\in]0;\infty[$
\begin{equation}
    b^\dagger(\xi)\alpha^\dagger(\eta)=\frac{1+2\eta}{1+2\eta+4\eta\xi}\alpha^\dagger(\eta+2\xi\eta)a\left(-\frac{\xi}{1+2\eta+2\xi+4\xi\eta}\right)
\end{equation}

$\bullet$ In the region $\eta\in]0;+\infty[$, $\xi\in]-\infty;-\frac{1}{2}[$
\begin{equation}
    \alpha(\xi)b(\eta)=\frac{1+2\xi}{1+2\xi+4\xi\eta}a^\dagger\left(-\frac{\eta}{1+2\eta+2\xi+4\xi\eta}\right)\alpha(\xi+2\eta\xi)
\end{equation}

$\bullet$ In the region $\eta\in]0;-\frac{1}{2}-\xi[$, $\xi\in]-\infty;-\frac{1}{2}[$
\begin{equation}
    \alpha(\xi)a^\dagger(\eta)=\frac{(1+2\eta)(1+2\xi)}{1+2\eta+2\xi}b\left(-\frac{\eta}{1+2\eta+2\xi}\right)\alpha\left(\frac{\xi}{1+2\eta}\right)
\end{equation}

$\bullet$ In the region $\eta\in]-\frac{1}{2}-\xi;\infty[$, $\xi\in]-\infty;-\frac{1}{2}[$
\begin{equation}
    \alpha(\xi)a^\dagger(\eta)=-\frac{(1+2\eta)(1+2\xi)}{1+2\eta+2\xi}\alpha^\dagger\left(\frac{\eta}{1+2\xi}\right)b^\dagger\left(-\frac{\xi}{1+2\xi+2\eta}\right)
\end{equation}

$\bullet$ In the region $\eta\in]-\frac{1}{2}-\xi;-\frac{1}{2}[$, $\xi\in]0;+\infty[$
\begin{equation}
    a(\xi)\alpha^\dagger(\eta)=-\frac{(1+2\eta)(1+2\xi)}{1+2\eta+2\xi}b\left(-\frac{\eta}{1+2\eta+2\xi}\right)\alpha\left(\frac{\xi}{1+2\eta}\right)
\end{equation}

$\bullet$ In the region $\eta\in]-\infty;-\frac{1}{2}-\xi[$, $\xi\in]0;+\infty[$
\begin{equation}
   a(\xi)\alpha^\dagger(\eta)=\frac{(1+2\eta)(1+2\xi)}{1+2\eta+2\xi}\alpha^\dagger\left(\frac{\eta}{1+2\xi}\right)b^\dagger\left(-\frac{\xi}{1+2\xi+2\eta}\right)
\end{equation}
\\
\\
From the $[\hat{\phi}(x_1),\hat{\phi}(x_2)]=0$ commutator, the resulting relations are:
\\
~
\\

$\bullet$ In the region $\eta\in]0;+\infty[$, $\xi\in]0;\infty[$
\begin{equation}
\label{eq:bdbd}
b^\dagger(\xi)b^\dagger(\eta)=b^\dagger\left(\eta+2\xi\eta\right)b^\dagger\left(\frac{\xi}{1+2\eta+4\xi\eta}\right)
\end{equation}
\begin{equation}
    a(\xi)b^\dagger(\eta)=b^\dagger\left(\frac{\eta}{1+2\xi}\right)a\left(\frac{\xi}{1+2\eta}\right)
\end{equation}
\begin{equation}
\label{eq:aa}
a(\xi)a(\eta)=a\left(\frac{\eta}{1+2\xi+4\xi\eta}\right)a\left(\xi+2\xi\eta\right)
\end{equation}

$\bullet$ In the region $\xi\in]0;+\infty[$, $\eta\in]0;\frac{1}{4\xi}[$
\begin{equation}
b^\dagger(\xi)a(\eta)=a\left(\frac{\eta+2\xi\eta}{1-4\xi\eta}\right)b^\dagger\left(\frac{\xi+2\xi\eta}{1-4\xi\eta}\right)
\end{equation}

$\bullet$ In the region $\xi\in]0;+\infty[$, $\eta\in]\frac{1}{4\xi};\infty[$
\begin{equation}
b^\dagger(\xi)a(\eta)=\alpha\left(\frac{\eta+2\xi\eta}{1-4\xi\eta}\right)\alpha\left(-\frac{\xi+2\xi\eta}{1+2\xi}\right)
\end{equation}

$\bullet$ In the region $\xi\in]0;+\infty[$, $\eta\in]-\infty;-\frac{1}{2}[$
\begin{equation}
b^\dagger(\xi)\alpha(\eta)=\alpha\left(\frac{\eta+2\xi\eta}{1-4\xi\eta}\right)a\left(-\frac{\xi+2\xi\eta}{1+2\xi}\right)
\end{equation}

$\bullet$ In the region $\xi\in]0;+\infty[$, $\eta\in]-\frac{1+2\xi}{4\xi};-\frac{1}{2}[$
\begin{equation}
a(\xi)\alpha(\eta)=\alpha\left(\frac{\eta}{1+2\xi+4\xi\eta}\right)b^\dagger\left(-\frac{\xi+2\xi\eta}{1+2\xi+4\xi\eta}\right)
\end{equation}

$\bullet$ In the region $\xi\in]0;+\infty[$, $\eta\in]-\infty;-\frac{1+2\xi}{4\xi}[$
\begin{equation}
a(\xi)\alpha(\eta)=a\left(\frac{\eta}{1+2\xi+4\xi\eta}\right)\alpha\left(\xi+2\eta\xi\right)
\end{equation}

$\bullet$ In the region $\xi\in]-\infty;-\frac{1}{2}[$, $\eta\in]0;-\frac{1}{2}-\xi[$
\begin{equation}
    \alpha(\xi)b^\dagger(\eta)=a\left(-\frac{\eta}{1+2\xi+2\eta}\right)\alpha\left(\frac{\xi}{1+2\eta}\right)
\end{equation}

$\bullet$ In the region $\xi\in]-\infty;-\frac{1}{2}[$, $\eta\in]-\frac{1}{2}-\xi;+\infty[$
\begin{equation}
    \alpha(\xi)b^\dagger(\eta)=\alpha\left(-\frac{\eta}{1+2\xi+2\eta}\right)b^\dagger\left(-\frac{\xi}{1+2\eta+2\xi}\right)
\end{equation}

$\bullet$ In the region $\xi\in]-\infty;-\frac{1}{2}[$, $\eta\in]0;+\infty[$
\begin{equation}
    \alpha(\xi)a(\eta)=b^\dagger\left(-\frac{\eta}{1+2\eta+2\xi+4\xi\eta}\right)\alpha(\xi+2\xi\eta)
\end{equation}

$\bullet$ In the region $\xi\in]-\infty;-\frac{1}{2}[$, $\eta\in]-\infty;-\frac{1}{2}[$
\begin{equation}
\label{eq:betabeta}
\alpha(\xi)\alpha(\eta)=b^\dagger\left(-\frac{\eta}{1+2\xi+2\eta+4\eta\xi}\right)a\left(\xi+2\eta\xi\right)
\end{equation}

Commutation relations for the $[\hat{\phi}^{\dagger}(x_1),\hat{\phi}^{\dagger}(x_2)]=0$ commutator can be obtained by taking the Hermitian conjugate of the commutators stemming from $[\hat{\phi}(x_1),\hat{\phi}(x_2)]=0$.

\subsection{Representation of the deformed oscillator algebra}

\begin{figure}[b!]
    \centering
    \includegraphics[scale=1]{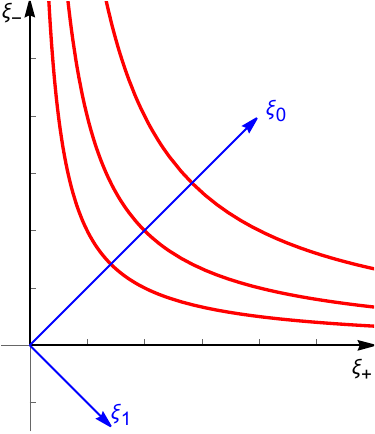}
    \caption{The mass-shells (in red) in light-cone coordinates are all confined in the $\xi_+ >0$, $\xi_- >0$ region.}
    \label{fig:mass_shells}
\end{figure}

A useful technique employed in studies of quantum field theories on non-commutative space-time is to represent the creation and annihilation operators of the deformed theory in terms of the ones of the standard theory \cite{Balachandran:2005eb,Joung:2007qv}. We introduce operators  $c$ and $c^\dagger$ which satisfy the standard bosonic commutation relations (in lightcone coordinates~\cite{Heinzl:2000ht}):
\begin{equation}
\label{eq:undefcommrels}
    [c(\xi),c^\dagger(\eta)]=2\xi\delta(\xi-\eta) \qquad [c(\xi),c(\eta)]=[c^\dagger(\xi),c^\dagger(\eta)]=0 \, ,
\end{equation}
for any real value of $\xi,\eta$. These operators act on the usual Fock space employed in quantum field theory. The vacuum state $\ket{0}$ is annihilated by $c(\xi)$ and $c^\dagger(-\xi)$, for $\xi>0$. Then, single particle states are defined as excitations of the vacuum state as 
\begin{equation}  
\label{eq:undefpart}
c^\dagger(\xi)\ket{0}=\ket{\xi}_{P} \qquad \xi>0  \,,
\end{equation}
while single anti-particle states are instead defined by
\begin{equation}
\label{eq:undefantipart}
    c(-\xi)\ket{0}=\ket{\xi}_{AP} \qquad \xi>0 \,.
\end{equation}
This parametrization of the standard oscillator algebra might not be familiar to the reader: it is a compact way of expressing the bosonic algebra of a complex scalar field in terms of a single infinite one-parameter set of operators. This is possible because, when expressed in lightcone coordinates, the creation and annihilation operators for particles and antiparticles depend on a single positive parameter, $\xi >0$, which is the lightcone momentum (\cref{fig:mass_shells}). Instead of having different symbols for the particle and antiparticle operators, we define $c(\xi)$ on negative values of $\xi$ too, and identify $c(\xi)$ for negative $\xi$ with the creation operator for the antiparticles. The corresponding annihilation operators will be the Hermitian conjugates of those. This choice just amounts to a relabeling of the Fourier coefficients of our scalar fields, and makes the notation more compact.

We recall the expressions for the Poincaré charges in terms of these operators.
The boost operator reads
\begin{equation}
\label{eq:caboost}
    N= -i\int_{0}^{\infty}\frac{d\xi}{2\xi}\xi\left(\frac{dc^\dagger(\xi)}{d\xi}c(\xi)+\frac{dc(-\xi)}{d\xi}c^\dagger(-\xi) \right) \, ,
\end{equation}
while the translations generators are given by 
\begin{equation}
\label{eq:catransl}
    P_+=\int_0^{\infty}\frac{d\xi}{2}(c^\dagger(\xi)c(\xi)+c(-\xi)c^\dagger(-\xi)), \qquad P_-=\int_0^{\infty}m^2\frac{d\xi}{2\xi^2}(c^\dagger(\xi)c(\xi)+c(-\xi)c^\dagger(-\xi)) \, . 
\end{equation}
These generators close the standard Poincaré algebra given that they are undeformed. 
Using \eqref{eq:undefcommrels} and \eqref{eq:caboost}, it is also easy to show that
\begin{equation}
\label{eq:boostcomms}
    [N,c(\xi)]=-i\xi\frac{dc(\xi)}{d\xi} \qquad [N,c^\dagger(\xi)]=-i\xi\frac{dc^\dagger(\xi)}{d\xi} \, , 
\end{equation}
for every $\xi$, and hence
\begin{equation}
\label{eq:boostaction}
\begin{aligned}
    &e^{ix N}c(\xi)e^{-ixN}=c(e^{x}\xi), \\
    &e^{ix N}c^\dagger(\xi)e^{-ixN}=c^\dagger(e^{x}\xi)  \, .
\end{aligned}
\end{equation}
For what follows, it is convenient to introduce the shorthand notation for the following finite boost transformation with momentum-dependent rapidity:
\begin{equation}
\label{eq:finiteboost}
    e^{i\ln(1+2\xi)N}\coloneqq B_\xi \, .
\end{equation}
It allows to write a representation for $a(\xi),b^\dagger(\xi)$ in a compact way, as follows (for $\xi>0$)
\begin{equation}
\begin{aligned}
\label{eq:representations1}
    &a(\xi)=\frac{1}{\sqrt{1+2S(\xi)}}B_{S(\xi)}c(\xi)=\frac{1}{\sqrt{1+2S(\xi)}}c(-S(\xi))B_{S(\xi)}\\
    &b^\dagger(\xi)=\frac{1}{\sqrt{1+2\xi}}c(-\xi)B_\xi=\frac{1}{\sqrt{1+2\xi}}B_\xi c(S(\xi))\\ 
    &a^\dagger(\xi)=\frac{1}{\sqrt{1+2S(\xi)}}c^\dagger(\xi)B_\xi=\frac{1}{\sqrt{1+2S(\xi)}}B_\xi c^\dagger(-S(\xi))\\
    &b(\xi)=\frac{1}{\sqrt{1+2\xi}}B_{S(\xi)}c^\dagger(-\xi)=\frac{1}{\sqrt{1+2\xi}}c^\dagger(S(\xi))B_{S(\xi)}  \, .
    \end{aligned}
\end{equation}
When $\xi<-1/2$, the commutators involving $\alpha(\xi),\alpha^\dagger(\xi)$ impose the constraints
\begin{equation}
\label{eq:repalpha}
\begin{aligned}
& \alpha(\xi)=\frac{1}{\sqrt{1+2S(\xi)}}B_{S(\xi)}c(\xi)=a(\xi)=b^\dagger(S(\xi))\\
&\alpha^\dagger(\xi)=\frac{1}{\sqrt{1+2S(\xi)}}c^\dagger(\xi)B_\xi=a^\dagger(\xi)=b(S(\xi))
\end{aligned}
\end{equation}
which can be obtained by extending the definitions in \eqref{eq:representations1} to negative $\xi$. Notice that these constraints also identify $a(\xi)$ with $b^\dagger(S(\xi))$, which is consistent with \eqref{eq:representations1} and with the commutative limit. 
Indeed, when $\kappa\rightarrow\infty$, the momentum space boundary $\xi=-\kappa/2$ vanishes, so that $a$ and $b^\dagger$ are not constrained anymore, as it should be in the commutative quantum field theory of the complex scalar field.

The presence of the particular operator $B_\xi$ in these representations is by no means incidental. It can be traced back to the plane wave flip governed by the $R$-matrix \eqref{eq:ospwflip}, which is explicitly dependent on $N$. When changing variables in the integrals appearing in the covariant quantization procedure, the \textit{braiding} of the momenta in plane waves is then reflected in the arguments of the creation and annihilation operators.   The deformed harmonic oscillator algebra~\eqref{eq:repalpha} is quite different from the one found in \cite{Fiore:2008sj} for $\theta$-Moyal non-commutative quantum field theory. There, the arguments of the creation and annihilation operators are left untouched, but the commutation relations are deformed by multiplication of a phase, dependent on $p_\mu\theta^{\mu\nu}q_\nu$, with $p,q$ being the momenta involved and $\theta^{\mu\nu}$ the antisymmetric matrix controlling the noncommutativity between coordinates.

Having represented the deformed creation and annihilation operators in terms of ordinary ones, we can now define (anti)-particle states of the deformed theory making use of the ordinary operators $c$, $c^\dagger$ on the standard Fock space.

\subsection{1-particle Fock state and C, P, T symmetries}
\label{subsec:Charge}

We start by exploring the $1$-particle states of the deformed theory. They are elements of the $1$-particle Hilbert space $\mathcal{H}$. From representations \eqref{eq:representations1}, it is immediate to see the the vacuum of the ordinary theory, $\ket{0}$ is also annihilated by the annihilation operators of the deformed theory 
\begin{equation}
    a(\xi)\ket{0}=b(\xi)\ket{0}=0 \, . 
\end{equation}
Single particle states are then defined as excitations of the vacuum 
\begin{equation}
\label{eq:1particle}
     \frac{a^\dagger(\xi)}{\sqrt{1+2\xi}}\ket{0}=c^\dagger(\xi)B_\xi\ket{0}=c^\dagger(\xi)\ket{0}=\ket{\xi}_P \, ,
\end{equation}
where we used the fact that $B_\xi\ket{0}=\ket{0}$ for every $\xi$ and the square root factor in the denominator guarantees normalization. Single anti-particle states are instead given by 
\begin{equation}
\label{eq:1antiparticle}
    \frac{b^\dagger(\xi)}{\sqrt{1+2S(\xi)}}\ket{0}=c(-\xi)B_\xi\ket{0}=c(-\xi)\ket{0}=\ket{\xi}_{AP} \, . 
\end{equation}
The single (anti)-particle states are thus identical to the ones defined in the commutative quantum field theory. As a consequence of this, the action of the momentum operators $P_\pm$ defined in \eqref{eq:catransl} give the standard results
\begin{equation}
    \label{eq:translact}
    \begin{aligned}
    &P_+\ket{\xi}_P=\xi \ket{\xi}_P \,, & \qquad  &P_+\ket{\xi}_{AP}=\xi \ket{\xi}_{AP} \,,\\
    &P_-\ket{\xi}_P=\frac{m^2}{\xi} \ket{\xi}_P \,,&   &P_-\ket{\xi}_{AP}=\frac{m^2}{\xi} \ket{\xi}_{AP} \,.
    \end{aligned}
\end{equation}
What about the $\alpha(\xi)$ operator? By letting it act on the vacuum, it is easy to see that $\alpha(\xi)\ket{0}=\alpha^\dagger(\xi)\ket{0}=0$. So we see that on the one-particle states, the $\alpha(\xi)$ leave no observable trace.

We attempt to define the charge conjugation operator as is ordinarily done in standard quantum field theory. We require that
\begin{equation}
\mathcal{C}\hat{\phi}(x^-,x^+)\mathcal{C}^{-1}=\hat{\phi}^\dagger(x^-,x^+) \, . 
\end{equation}
Recalling the expression for the scalar field \eqref{eq:sfexpr}, the above constraint yields, for $\xi>0$
\begin{equation}
\label{eq:ccstd}
    \mathcal{C}\frac{a(\xi)}{1+2\xi}\mathcal{C}^{-1}=b(\xi) \qquad \mathcal{C}\, b^\dagger(\xi)\, \mathcal{C}^{-1}=\frac{a^\dagger(\xi)}{1+2\xi} \, , 
\end{equation}
while for $\xi<-1/2$, we have
\begin{equation}
\label{eq:ccalpha}
    \mathcal{C}\frac{\alpha(\xi)}{1+2\xi}\mathcal{C}^{-1}=-\alpha^\dagger(S(\xi)) \, . 
\end{equation}
For single particle states, \eqref{eq:ccstd} yields simply
\begin{equation}
    \mathcal{C}\ket{\xi}_{AP}=\ket{\xi}_P \, ,
\end{equation}
as is the case in the undeformed quantum field theory. As a result, we can write the charge conjugation operator as

\begin{equation}
    \label{chargeconjugation}\mathcal{C}=\int_0^{\infty}d\xi \qty[c^\dagger(\xi)c^\dagger(-\xi)+c(\xi)c(-\xi)] \, ,
\end{equation}
which is just the usual expression one obtains also in commutative quantum field theory. Using the above and representations~\eqref{eq:representations1}, \eqref{eq:repalpha} for the creation and annihilation operators, properties \eqref{eq:ccstd} and \eqref{eq:ccalpha} can be explicitly verified.

A remark on the consequences of Eq.~\eqref{chargeconjugation}: as can be seen from Eq.~\eqref{eq:translact}, the one-particle state and the one-antiparticle state associated to it through the charge conjugation operator carry the same momentum. This departs from what was recently found in the timelike $\kappa$-Minkowski case in~\cite{Arzano:2020jro}, where it appears that the charge conjugation operator sends a one-particle state into a one-antiparticle state with different momentum. This led to an interesting phenomenology and the possibility of putting bounds to the noncommutativity parameters only a few order of magnitude lower than the Planck energy~\cite{Arzano:2019toz,Arzano:2020rzu}. Unfortunately, these experimental bounds are irrelevant for the model considered in this paper.

Regarding parity (P) and time revesal (T), \textbf{in the commutative case, in lightcone coordinates,} these operators are introduced as, respectively:
\begin{equation}
    P:\,  x^\pm \to x^\mp  \,, \qquad     T: \,  x^\pm  \to - x^\mp \,,
\end{equation}
which are mapped to two involutive operators $\mathcal{P}$ and $\mathcal{T}$ acting on the creation and annihilation operators, defined by
\begin{equation}\label{eq:PTactingonFields}
\mathcal{P} \hat \phi(x^-,x^+)\mathcal{P}^{-1}  =\hat \phi(x^+,x^-) \,, \qquad
 \mathcal{T}\hat \phi(x^-,x^+)\mathcal{T}^{-1}  =\overline{\hat \phi(-x^+,-x^-) }\,,
\end{equation}
where the operator over the quantum field on the right hand side of the action of the $\mathcal{T}$ operator is a complex conjugate, as opposed to a Hermitian conjugate, as it acts only on the plane waves in the Fourier expansion of the fields, and leaves the construction and annihilation operators unchanged. It is necessary to compose the na\"ive time reversal operator with this complex conjugate, thereby making the operator \textit{antilinear,} in order to have a well-behaved transformation on the Fock space (the na\"ive operator $\hat \phi(x^-,x^+) \to \hat \phi(-x^+,-x^-)$ would be unacceptable, as it would end up annihilating all one-particle states~\cite{Peskin:1995ev}).  Replacing in the above the expansion of an on-shell quantum field in creation and annihilation operators [\textit{i.e.} the commutative equivalent of Eq.~\eqref{eq:sfexpr}], one gets the following action of $\mathcal{P}$ and $\mathcal{T}$:
\begin{equation}\label{eq:PTactingonOscillators}
\begin{aligned}
        &\mathcal{P}a(\xi)\mathcal{P}^{-1} =  \pm a\qty(\frac{m^2}{\xi})
    \,,&
    &\mathcal{P}b(\xi)\mathcal{P}^{-1}  =   \pm b\qty(\frac{m^2}{\xi})
    \,, \\
        &\mathcal{P}a^\dagger(\xi)\mathcal{P}^{-1}  =  \pm  a^\dagger\qty(\frac{m^2}{\xi})
    \,,&
    &\mathcal{P}b^\dagger(\xi_+)\mathcal{P}^{-1}  =  \pm  b^\dagger\qty(\frac{m^2}{\xi}) \,,
\end{aligned}
\end{equation}
and
\begin{equation}
\begin{aligned}
        &\mathcal{T}a(\xi)\mathcal{T}^{-1}  = a\qty(\frac{m^2}{\xi})
    \,,&
    &\mathcal{T}b(\xi)\mathcal{T}^{-1} = b\qty(\frac{m^2}{\xi})
    \,, \\
        &\mathcal{T}a^\dagger(\xi)\mathcal{T}^{-1} = a^\dagger\qty(\frac{m^2}{\xi})
    \,,&
    &\mathcal{T}b^\dagger(\xi)\mathcal{T}^{-1}  = b^\dagger\qty(\frac{m^2}{\xi}) \,.
\end{aligned}
\end{equation}
Acting on the vacuum with the left- and right-hand sides of the equations above, one gets:
\begin{equation}
\begin{aligned}
     &\mathcal{P}\ket{\xi}_{P} = \pm  \ket{\frac{m^2}{\xi}}_{P} \,, &  \
     &\mathcal{P}\ket{\xi}_{AP} = \pm  \ket{\frac{m^2}{\xi}}_{AP} &\,,
     \\
    &\mathcal{T}\ket{\xi}_P =   \ket{\frac{m^2}{\xi}}_P  \,,&
     &\mathcal{T}\ket{\xi}_{AP} =  \ket{\frac{m^2}{\xi}}_{AP} \,,
\end{aligned}
\end{equation}
where the $\pm$ sign depends on the parity of the particle and the overall phase for the time-reversal was omitted since it has no effect on our discussion.

One could imagine to extend this analysis to the noncommutative case, exactly like what we did in Subsec.~\ref{subsec:Charge} for the charge conjugation operator. However, an obstacle immediately manifests itself: there is no sense in which the coordinate commutation relations~\eqref{eq:cnc} can be invariant under parity and time-reversal transformations. In the noncommutative case, we have to choose what these operators do to the noncommutative product between coordinates: they may leave it unchanged, meaning that they are \textit{homomorphisms} for this product, or they may exchange the product order, in which case they are \textit{anti-homomorphisms.} This distinction is absent in the commutative case, precisely because the products are commutative. So, for consistency with the commutative limit, we need a linear P operator and an antilinear T operator, however we are free to choose either of them as homomorphisms or anti-homomorphisms. Regardless of what we choose, since the commutation relations~\eqref{eq:cnc} have $x^-$ on the right-hand side, an operator that sends $x^-$ to $x^+$ can never leave them invariant.

If we insist on introducing a P and a T operator as in~\eqref{eq:PTactingonFields}, acting on our on-shell noncommutative quantum fields~\eqref{eq:sfexpr}, then our on-shell plane waves are sent to off-shell ones. For example, choosing P to be a homomorphism one gets the following transformation rule for a noncommutative plane wave:
\begin{equation}
     e[\xi] \to   e^{i\frac{m^2}{\xi}x^{+}} e^{\frac{i}{2}\ln(1+2\xi)x^-} = e^{\frac{i}{2}\ln(1+2\xi) e^{-2 \, \frac{m^2}{\xi} }x^-} e^{i\frac{m^2}{\xi}x^{+}} \,,
\end{equation}
and the pair of momentum components that appear on the right hand side:
\begin{equation}
    \left(\frac{1}{2}\ln(1+2\xi) e^{-2 \, \frac{m^2}{\xi} } , \frac{m^2}{\xi} \right)  \, ,
\end{equation}
does not satisfy the on-shell relation anymore. The same happens for the other on-shell waves in the field expansion, including those of ``new type''. If we chose P to be an anti-homomorphism:
\begin{equation}
    e[\xi] \to   e^{\frac{i}{2}\ln(1+2\xi)x^-}  e^{i\frac{m^2}{\xi}x^{+}}  \,,
\end{equation}
we end up with the following  a pair of momentum components:
\begin{equation}
    \left(\frac{1}{2}\ln(1+2\xi) , \frac{m^2}{\xi} \right) \,,
\end{equation}
which again does not satisfy the on-shell relation (the $\xi_-$ and $\xi_+$ components are in the wrong order). Analogous calculations can also be done for the time reversal operator $\mathcal{T}$, and still result in an off-shell plane wave.

This is just a manifestation of the  non-invariance of the basic  commutation relations~\eqref{eq:cnc}, which are the starting point of the whole model. This theory is parity- and time-reversal-breaking. 
However, the theory can still be said to preserve combined PT invariance: if both P and T are chosen to have the same behaviour with respect to the noncommutative product, \textit{i.e.} they are both homomorphisms or anti-homomorphisms, and P is assumed linear while T is assumed antilinear, then the  coordinate commutation relations~\eqref{eq:cnc} turn our to be invariant. Such a PT operator would leave also the on-shell plane waves $e[\xi]$ invariant, however the new-type waves could, in principle, change: looking at Eq.~\eqref{eq:newpws}, $\mathcal{E}_a[\xi] =  E_a[\xi]e^{-\frac{\pi}{2}x_a^+}$, it is clear that, an antilinear homomorphism like our PT operator would leave $ E_a[\xi]$ invariant, while changing the $e^{-\frac{\pi}{2}x_a^+}$ term into $e^{+\frac{\pi}{2}x_a^+}$. This, however, is harmless, as we can easily prove that $e^{-\frac{\pi}{2}x_a^+} = e^{+\frac{\pi}{2}x_a^+}$ in our representation, when acting on functions of a single variable. Thus,  the new-type plane waves are also left invariant by PT.

It appears that PT transformations are still a symmetry of our theory. In particular, PT acts like the identity on the scalar field Fock space (this is true in the commutative case too for spin-zero fields~\cite{Peskin:1995ev}).  Finally, CPT is preserved too.

\subsection{Braided flip operator and multiparticle states}
\label{sec: multiparticle states}

We now begin exploring the multi-particle sector of the theory. To get a feeling of the novelties introduced by our non-commutative framework, let us start by focusing on two particle states (the conclusions drawn will be analogous for anti-particle states): 
\begin{equation}
\label{eq:2part}
\begin{aligned}
\ket{\xi}_P\otimes \ket{\eta}_P = \frac{a^\dagger(\xi)}{\sqrt{1+2\xi}}\ket{0}\otimes \frac{a^\dagger(\eta)}{\sqrt{1+2\eta}}
\end{aligned}\ket{0} \, , 
\end{equation}
which are elements of the tensor product of two copies of the 1-particle Hilbert space $\mathcal{H}$.
The total momentum for this two-particle state is obtained by acting with the coproducts of the translation generators, as dictated by Hopf Algebra axioms when acting on the tensor product of its representation. To obtain the total $+$ component of the momentum, we apply the coproduct \eqref{eq:lincop} for $P_+$, 
\begin{equation}
\begin{aligned}
\label{eq:ppcopact}
    &\Delta [P_+] \qty(\ket{\xi }_P \otimes \ket{\eta}_P)= P_+\ket{\xi }_P\otimes \ket{\eta }_P+ \ket{\xi }_P \otimes P_+ \ket{\eta }_P + 2 P_+ \ket{\xi }_P\otimes P_+ \ket{\eta }_P=\\
    =& \qty(\xi+\eta+2\xi\eta) \qty(\ket{\xi }_P \otimes \ket{\eta}_P) = \Delta[\xi,\eta]_+\qty(\ket{\xi }_P \otimes \ket{\eta}_P) \,,
    \end{aligned}
\end{equation}
where the $\Delta[\xi,\eta]$ operation for liner momentum was defined in \eqref{eq:linoplus}.
In a similar fashion, we can calculate the $-$ component for the total momentum, yielding  
\begin{equation}
\begin{aligned}
\label{eq:pmcopact}
    &\Delta[P_-] \qty(\ket{\xi }_P \otimes \ket{\eta}_P)= P_-\ket{\xi }_P\otimes \ket{\eta }_P+ \ket{\xi }_P \otimes P_- \ket{\eta }_P -\frac{2 P_+}{1+2P_+} \ket{\xi }_P\otimes P_- \ket{\eta }_P=\\
    =& \qty (\frac{m^2}{\xi}+\frac{m^2}{\eta}-\frac{2\xi}{1+2\xi}\frac{m^2}{\eta})(\ket{\xi}_P\otimes \ket{\eta}_P) = \Delta[\xi,\eta]_-\qty(\ket{\xi }_P \otimes \ket{\eta}_P) \,,
    \end{aligned}
\end{equation}
The same line of reasoning can be applied to anti-particle states, obtaining the same results for the total momenta. 

In ordinary quantum field theory, multi-particle states live in symmetrized or anti-symmetrized tensor-products of single-particle states, which characterize the notion of identical particles. The key ingredient is the ordinary flip operator $\flip$, which is an involutive operation on the tensor product of Hilbert spaces of single particle states, where the multi-particle states are defined. In general, an analogous construction of the multi-particle Fock space is not so straightforward for quantum field theories on non-commutative space-time \cite{Arzano:2022vmh}. The main reason for this is that the standard flip operation applied to a two-particle state yields another two-particle state carrying different total momentum, due to the non-commutative nature of the coproducts.
In our specific model, this simply follows from observing that 
\begin{equation}
    \Delta [P_-] \qty(\ket{\xi }_P \otimes \ket{\eta}_P)\neq \Delta [P_-] \qty(\ket{\eta }_P \otimes \ket{\xi}_P) \,.
\end{equation} 

The way out of this \textit{empasse} is to define a deformed notion of particle exchange. This is possible, for example, in non-commutative quantum field theory on the $\theta$-Moyal non-commutative space-time \cite{Fiore:2007vg,Balachandran:2005eb}, due to the properties of the twist operator, linked to the existence of an $R$-matrix \cite{Aschieri:2007sq}. For the much-studied timelike $\kappa$-Minkowski case, several works \cite{Arzano:2008bt,Govindarajan:2009wt,Arzano:2022vmh} have tried to identify a \textit{braiding} of single-particle states in order to construct a deformed notion of symmetric and anti-symmetric states. 
These approaches all present some shortcomings: either the braiding is not involutive, or it is not covariant when constructing the theory at all orders in $\kappa$. The recent~\cite{Arzano:2022vmh} finds that, accepting a non-involutive flip operator as the physical one, the notion of identical particles has to be abandoned. The lack of involutivity of the flip operator leads, in fact, to an infinite tower of states characterized by the same total momentum. In the present work, we find that the $\kappa$-lightlike framework, although characterized by the same non-abelian momentum Lie-group structure as the timelike case, admits a well-defined notion of identical particles, thanks to the existence of the universal $R$-matrix. 
Consider, for instance, the two-particle state defined as 
\begin{equation}
\label{eq:simmstate}
\Tilde{R}\qty(\ket{\xi}_P\otimes \ket{\eta}_P)=R\circ \flip ~ \qty(\ket{\xi}_P\otimes \ket{\eta}_P)=R\qty(\ket{\eta}_P\otimes \ket{\xi}_P) \, ,
\end{equation}
\textit{i.e.}, we act with the flip operator $\flip$, where $\flip ~ \qty(\ket{\xi}_P\otimes \ket{\eta}_P) =  \ket{\eta}_P\otimes \ket{\xi}_P$, and then with the $R$-matrix defined in \eqref{eq:rmatfinal}.
In detail, we have 
\begin{equation}
\label{eq:Rflip}
    \begin{aligned}
         \Tilde{R}\qty( \ket{\xi}_P\otimes \ket{\eta}_{P})=&e^{-2i \ln(1+2P_+)\otimes N}e^{2i N\otimes \ln(1+2P_+)} (\ket{\eta}_P\otimes \ket{\xi}_P)=\\
          =&e^{-2i \ln(1+2P_+)\otimes N}\left(B_\xi \ket{\eta}_P\otimes \ket{\xi}_P\right)=\\
          =&\ket{\eta+2\xi\eta}_P\otimes B_{S(\eta+2\xi\eta)}\ket{\xi}_P=\\
          =&\ket{\eta+2\eta\xi}_P\otimes\ket{\frac{\xi}{1+2\eta+4\eta\xi}}_P \, .
    \end{aligned}
\end{equation}
The structure of the new obtained two-particle states mimicks the structure of the Hermitian conjugate of commutation relation \eqref{eq:aa}, where the deformation emerges from applying the $R$-matrix to exchange plane waves upon performing covariant quantization, as discussed in \cref{sec:covquant}.
By acting with the momentum coproducts \eqref{eq:lincop}, it is now easy to check that
\begin{equation}
    \Delta [P_{\pm}] \qty(\ket{\eta+2\eta\xi}_P\otimes\ket{\frac{\xi}{1+2\eta+4\eta\xi}}_P)=\Delta(\xi,\eta)_{\pm}\qty(\ket{\eta+2\eta\xi}_P\otimes\ket{\frac{\xi}{1+2\eta+4\eta\xi}}_P) \, ,
\end{equation}
so the deformed symmetric state \eqref{eq:simmstate} has the same total momentum as \eqref{eq:2part}, thus being a suitable candidate for our construction of deformed (anti-)symmetric states.
It is also easy to show that $\tilde{R}$ is an involutive operator, \textit{i.e.} $(\Tilde{R}^2=1)$. Indeed, repeating the same analysis as in \eqref{eq:Rflip}, one can show that 
\begin{equation}
    \Tilde{R}\, \left( \ket{\eta+2\eta\xi}_P\otimes\ket{\frac{\xi}{1+2\eta+4\eta\xi}}_P \right) = \ket{\xi}_P\otimes \ket{\eta}_{P} \, .
\end{equation}
This last property makes $\Tilde{R}$ and ideal candidate for constructing a deformed symmetrization operator, useful in defining deformed symmetric states in our field theory. We are now ready to define the deformed symmetrization operator:
\begin{equation}
\label{eq:symmop}
    \mathcal{S}^+\coloneqq \frac{1}{2}( 1\otimes 1+\Tilde{R}) \,, 
\end{equation}
which is such that $(\mathcal{S}^+)^2=\mathcal{S}^+$, \textit{i.e.}  $\mathcal{S}^+$ is idempotent. Then, we can define deformed symmetric two-particle states simply as 
\begin{equation}
\label{eq:braiding}\sqrt{2} \,\mathcal{S}^+\left(\ket{\xi}_P\otimes \ket{\eta}_P\right) \,, 
\end{equation}
where the $\sqrt{2}$ factor is introduced for normalization.
In an analogous way, we can define the antisymmetrization operator $\mathcal{S}^-$:
\begin{equation}
    \mathcal{S}^-\coloneqq \frac{1}{2}( 1\otimes 1-\Tilde{R}) \,, 
\end{equation}
which is also idempotent and can be used to define antisymmetric multi-particle states.\footnote{Although so far we only worked out the quantization of a scalar field, we can already say something about fermionic fields and their deformed Fock space, just by analyzing the general properties of the R-matrix.}

So far, we have shown that there exists an involutive braiding that suggests the definition of deformed symmetric two-particle states in lightlike $\kappa$-Minkowski quantum field theory. We now show the covariance of such braiding, in order to complete the picture.

A single particle state transforms under a finite boost of parameter $\tau$ as 
\begin{equation}
    \ket{\xi}_P\rightarrow e^{i\tau N}\ket{\xi}_P=\ket{e^{\tau}\xi}_P
\end{equation}
When acting on the tensor product of single-particle states,  the boost coproduct \eqref{eq:lincop} needs to be taken into account. For a finite transformation, using the commutation relations~\eqref{eq:lincomm}, we can prove that
\begin{equation}
\label{eq:finiteboostcop}
    e^{i\tau \, \Delta [N]}=e^{i\tau N}\otimes e^{i\ln(\frac{(1+2P_+)e^\tau}{1+2e^\tau P_+})N} \, .
\end{equation}
For our two particle state \eqref{eq:2part}, this yields 
\begin{equation}
    \ket{\xi}_P\otimes \ket{\eta}_P \rightarrow e^{i\tau N}\ket{\xi}_P\otimes e^{i\ln(\frac{(1+2\xi)e^\tau}{1+2\xi e^\tau})N}\ket{\eta}_P=\ket{e^{\tau}\xi}_P\otimes \ket{\frac{(1+2\xi)}{1+2e^\tau\xi}e^\tau\eta}_P\, . 
\end{equation}
The deformed flipped state \eqref{eq:Rflip} is instead mapped into 
\begin{equation}
\label{eq:flipboost}
    \ket{\eta+2\eta\xi}_P\otimes\ket{\frac{\xi}{1+2\eta+4\eta\xi}}_P \rightarrow \ket{e^\tau \eta(1+2\xi)}_P\otimes \ket{\frac{e^\tau\xi}{1+2e^\tau\eta(1+2\xi)}}_P \, . 
\end{equation}
Conversely, by first boosting the two-particle state \eqref{eq:2part} and then flipping it with $\Tilde{R}$, the result is 
\begin{equation}
    \label{eq:boostflip}\Tilde{R}\qty(\ket{e^{\tau}\xi}_P\otimes \ket{\frac{(1+2\xi)}{1+2e^\tau\xi}e^\tau\eta}_P)=\ket{e^\tau \eta(1+2\xi)}_P\otimes \ket{\frac{e^\tau\xi}{1+2e^\tau\eta(1+2\xi)}}_P \, ,
\end{equation}
which is identical to the right-hand side of \eqref{eq:flipboost}. We have thus proved that
\begin{equation}
    \tilde{R} e^{i\tau \Delta[N]}=e^{i\tau \Delta[N]}\tilde{R} \, .
\end{equation}
Basically, our deformed flip operator $\Tilde{R}$ commutes with all the Hopf Algebra generators $P_\pm,N$, also given its compatibility with the momenta coproducts shown above. Therefore, relativistic covariance is guaranteed. 

\section{Physical interpretation of deformed multi-particle states}
\subsection{On the indistinguishability of identical particles}
In quantum mechanics, two particles of the same species are described by a symmetric or anti-symmetric state \cite{sakurai_napolitano_2017}, defined as 
\begin{equation}
\label{eq:identicalpart}
    \sqrt{2}\qty(\frac{1\pm \flip}{2}) \ket{p}\otimes\ket{q}=\frac{\ket{p}\otimes \ket{q}\pm\ket{q}\otimes \ket{p}}{\sqrt{2}} \, ,
\end{equation}
where $\flip$ is the standard flip operator and $p,q$ are the linear momenta of the particles. Operationally, the indistinguishability of the two particles may be understood as follows. Suppose we have a calorimeter that can measure the energy of one particle at a time, from which we can deduce the corresponding momentum (we are in 1+1 dimensions). According to state \eqref{eq:identicalpart}, the calorimeter can measure either $p$ or $q$. If our calorimeter measures momentum $p$, for example, we have no way of knowing if the measured particle is the one in the first or second place of the tensor product. This indistinguishability simply follows from the (anti)-symmetric property of the quantum mechanical state describing the two-particle system. What happens then if the two-particle state is instead defined by the deformed (anti)-symmetrization operators $\mathcal{S}^{\pm}$ in \eqref{eq:symmop}? We reintroduce the dimensional parameter $\kappa$, for clarity. Consider a decay process of an initial particle of mass $M$ with momentum $\Pi_\mu = (\Pi , M^2/\Pi)$ (in light-cone coordinates). The particle decays into two identical particles of mass $m$ and momenta $\xi_\mu = (\xi , m^2/\xi)$,  $\eta_\mu = (\eta , m^2/\eta)$. We will call $\xi$ the momentum of the particle that enters the coproduct~\eqref{eq:linoplus}, in the deformed momentum conservation law, from the left, while $\eta$ is the label of the momentum on the right-hand side of the coproduct. Notice that this labeling choice has nothing to do with the placement of the particle momenta in the tensor product, and has no physical consequences: we could choose the opposite convention and nothing would change in the calculations. The deformed momentum conservation law dictates:
\begin{equation}
\left\{\begin{aligned}
    &
    \Pi = \xi + \eta + 2 \xi \, \eta \,,
    \\
    &
    \frac{M^2}{\Pi} = \frac{m^2}{\xi} + \left( 1  - \frac{2 \, \xi}{\kappa+2 \xi} \right) \frac{m^2}{\eta} \,,
\end{aligned}
\right.
\end{equation}
the above two equations can be solved with respect to $\xi$ and $\eta$, and they have two solutions (recall that all the on-shell momenta, $\Pi$, $\xi$ and $\eta$ are positive-definite):
\begin{equation}
    \begin{aligned}
        &\xi = F_1(\Pi,M,m) \,, &  & \eta = G_1(\Pi,M,m,\kappa) \,,
        \\
        &\xi = F_2(\Pi,M,m) \,, &  & \eta = G_2(\Pi,M,m,\kappa) \,,
    \end{aligned}
\end{equation}
where
\begin{equation}
\begin{gathered}
        F_1(\Pi,M,m) = \frac{\Pi }{2} \left( 1 + \sqrt{1 - \frac{4 m^2}{M^2}} \right) \,, \qquad     F_2(\Pi,M,m) = \frac{\Pi }{2} \left( 1 - \sqrt{1 - \frac{4 m^2}{M^2}} \right) \,,
    \\
 G_1(\Pi,M,m,\kappa) = \frac{\kappa  \Pi \left(M (\kappa +2 \Pi) \sqrt{M^2-4 m^2}-4 m^2 \Pi+M^2 (\kappa +2 \Pi)\right)}{8 m^2 \Pi^2+2 \kappa  M^2 (\kappa +2 \Pi)} \,,
\\
G_2(\Pi,M,m,\kappa)= \frac{\kappa  \Pi \left(-M (\kappa +2 \Pi) \sqrt{M^2-4 m^2}-4 m^2 \Pi+M^2 (\kappa +2 \Pi)\right)}{8 m^2 \Pi^2+2 \kappa  M^2 (\kappa +2 \Pi)} \,.
\end{gathered}
\end{equation}
If we choose the first solution, the final state will be the following:
\begin{equation}
\ket{\psi_1} = \sqrt{2} \,  \mathcal{S}^+ \big{(} \ket{F_1} \otimes \ket{ G_1} \big{)} = \frac{1}{\sqrt{2}} \bigg{(} \ket{F_1} \otimes \ket{ G_1} +   \tilde R \left[\ket{F_1} \otimes \ket{ G_1} \right] \bigg{)} \, ,
\end{equation}
while if we choose the second:
\begin{equation}
\ket{\psi_2} =  \sqrt{2} \,  \mathcal{S}^+ \big{(} \ket{F_2} \otimes \ket{ G_2} \big{)} = \frac{1}{\sqrt{2}} \bigg{(} \ket{F_2} \otimes \ket{ G_2} +   \tilde R \left[\ket{F_2} \otimes \ket{ G_2} \right] \bigg{)} \,.
\end{equation}
However, as it turns out, the two states are identical. In fact, it is possible to show that
\begin{equation}
  \tilde R \left[\ket{F_1} \otimes \ket{ G_1} \right] = \ket{F_2} \otimes \ket{ G_2} \,,
  \qquad
  \tilde R \left[\ket{F_2} \otimes \ket{ G_2} \right] = \ket{F_1} \otimes \ket{ G_1} \,,     
\end{equation}
implying $\ket{\psi_1} = \ket{\psi_2} $, just like in the undeformed theory.
 This is a nontrivial rigidity of the theory, consequence of the Hopf-algebraic constraints that entail its relativistic nature. The compatibility between the deformed momentum composition law and the flip operator is what is behind it.
The final state, $\ket{\psi_1} = \ket{\psi_2} $ is proportional the sum of the following two kets (at first order in $\kappa^{-1}$):
\begin{equation}
\begin{aligned}
    &\ket{F_1} \otimes \ket{ G_1}  &= \ket{F_1}\otimes\ket{F_2-\frac{2 m^2 \Pi^2}{\kappa \, M^2} + \mathcal{O}(\kappa^{-2})} \,,
\\
&\tilde R \left[\ket{F_1} \otimes \ket{ G_1} \right]  &= \ket{F_2}\otimes\ket{F_1 -\frac{2 m^2 \Pi^2}{\kappa \, M^2} + \mathcal{O}(\kappa^{-2})} \,.
\end{aligned}
\end{equation}
The result is consistent with the commutative limit $\kappa\rightarrow\infty$. However, when the $\kappa$-deformation is switched on, the qualitative features of this multi-particle state are completely different from their undeformed counterpart.
If the calorimeter measures the momentum of one of the particles, we can obtain one of the following four results:
\begin{equation}\label{eq:FinalFourMomenta}
    \begin{aligned}
&\text{Unflipped, Left:}& &\frac{\Pi }{2} \left( 1 + \sqrt{1 - \frac{4 m^2}{M^2}} \right)\,,
        \\
&\text{Flipped, Left:}& &\frac{\Pi }{2} \left( 1 - \sqrt{1 - \frac{4 m^2}{M^2}} \right)\,,
\\
&\text{Flipped, Right:}& &\frac{\Pi }{2} \left( 1 + \sqrt{1 - \frac{4 m^2}{M^2}} \right)
-\frac{2 m^2 \Pi^2}{\kappa \, M^2} + \mathcal{O}(\kappa^{-2})\,,
\\
&\text{Unflipped, Right:}& &\frac{\Pi }{2} \left( 1 - \sqrt{1 - \frac{4 m^2}{M^2}} \right)
-\frac{2 m^2 \Pi^2}{\kappa \, M^2} + \mathcal{O}(\kappa^{-2}) \,,
    \end{aligned}
\end{equation}
according to whether we are measuring the left- or right-hand side of the tensor product of the unflipped state, $\ket{F_1} \otimes \ket{ G_1} $, or of the flipped state, $\tilde R \left[\ket{F_1} \otimes \ket{ G_1} \right] $.
In the noncommutative theory, the four momenta~\eqref{eq:FinalFourMomenta} are all different. Measuring the momentum of one particle allows us to identify which side of the tensor product it came from, and from which state (flipped or unflipped). Therefore, there is a sense in which the indistinguishability of identical particles is lost when the $\kappa$-deformation is taken into account. There is no avoiding this if we want to construct a relativistic theory. As already stressed in \cref{sec: multiparticle states}, a state of the type \eqref{eq:identicalpart} would not be covariant under the $\kappa$-Poincaré transformations, which exhibit all their non-trivial behaviour on multi-particle states, given that the coproduct is involved.

\subsection{On the Pauli exclusion principle}
\begin{figure}
    \centering
    \includegraphics[scale=0.8]{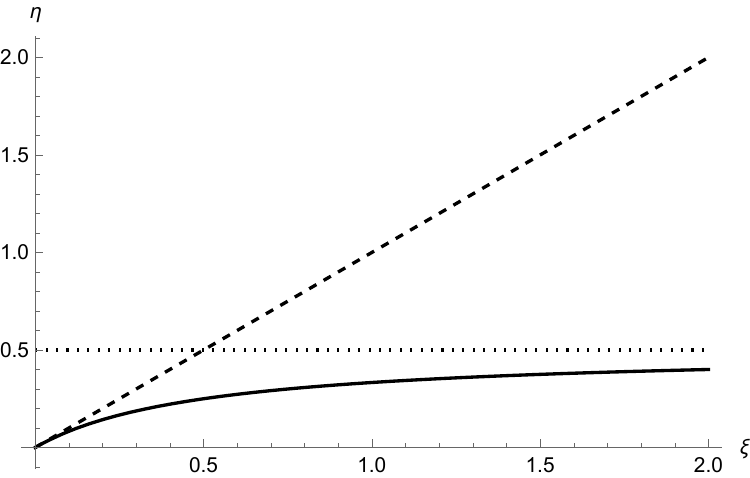}
    \caption{The $+$-momentum space of momentum-pairs for two-particle states, in units of $\kappa$. The dashed line represents the pairs excluded by the undeformed Pauli exclusion principle. The thick curve represents the pairs excluded by the deformed version of the exclusion principle when non-commutativity is taken into account. For large $\xi$, the curve saturates at $1/2\kappa$, which is the dotted asymptote in the plot.}
    \label{fig:pauli}
\end{figure}

We now explore the consequences of the deformed permutation symmetry on the Pauli exclusion principle. In standard quantum theory, a state describing two fermions with the same quantum numbers is annihilated by the undeformed anti-symmetrizer: 
\begin{equation}
\label{eq:idferm}
    \qty(\frac{1-\flip}{2})\ket{\xi}\otimes\ket{\xi}=0 \, . 
\end{equation}
This is the essence of the Pauli exclusion principle, which has been confirmed in a variety of experiments searching for classically prohibited transitions to states of the form $\ket{\xi}\otimes\ket{\xi}$. It is then natural to ask what is the fate of the Pauli exclusion principle in our $\kappa$-deformed framework. Assuming that 2-fermion states are left invariant by $\mathcal{S}^-$, what is the class of states annihilated by this operator? We require that:
\begin{equation}
    \mathcal{S}^-\qty(\ket{\xi}_P\otimes \ket{\eta}_P)=0 \, ,
\end{equation}
which uniquely selects the two particle states 
\begin{equation}\label{eq:DeformedIdenticalParticles}
 \ket{\xi}_P \otimes \ket{-S(\xi)}_P = 
 \ket{\xi}_P \otimes \ket{\xi/(1+2\xi)}_P \,.  
\end{equation}
The same holds for $\ket{\eta/(1-2\eta)} \otimes \ket{\eta}$, which is the same state as~\eqref{eq:DeformedIdenticalParticles}, just parametrized with respect to the momentum of the particle on the right hand side of the tensor product. In the undeformed case, the solution $\xi=\eta$ would have been selected, in agreement with \eqref{eq:idferm}. In light of this reasoning, we can visualize the Pauli principle in a simple manner. In the $(\xi,\eta)$ plane, which contains admissible pairs of momenta that can be attributed to fermions, the Pauli principle excludes a one-dimensional subset of the $(\xi,\eta)$ plane: the pairs lying on a curve $\eta=f(\xi)$. In the commutative case, $f(\xi)=\xi$ (the bisector of the plane), while the $\kappa$-deformed version is $f(\xi)=-S(\xi)$ (see \cref{fig:pauli}) . 

Notice that our deformed identical-particles states are Lorentz-covariant,
\begin{equation}
    \Delta(e^{i\tau N})  \ket{\xi}_P \otimes \ket{-S(\xi)}_P =  \ket{e^\tau \xi}_P \otimes \ket{-S( e^\tau \xi)}_P \,,
\end{equation}
in the sense that identical-particles states are sent to boosted identical-particles states by a Lorentz transformation. This is due to the fact that the curve $\eta = - S(\xi)$ lays on an orbit of the Lorentz group.

In light of the previous observations, we notice that the state $\ket{\xi}_P\otimes\ket{\xi}_P$ is not annihilated by $\mathcal{S}^-$, contrary to the commutative case. Notice, however, that this form of the state is not preserved by Lorentz transformations: if an observer attributes the same momentum $\xi$ to two particles, by the action of  the finite boost generator~\eqref{eq:finiteboostcop} on the state $\ket{\xi}_P\otimes\ket{\xi}_P$, a boosted observer would attribute different momenta to them:
\begin{equation}
\ket{\xi}_P\otimes\ket{\xi}_P \to \ket{e^\tau\xi}_P\otimes\ket{\frac{e^\tau\xi(1+2\xi)}{1+2e^\tau\xi}}_P  \,, 
\end{equation}
with $\tau$ being the boost parameter.

The discussion above highlights the fact that, in our model, states of the form $\ket{\xi}_P\otimes\ket{\xi}_P$  are not excluded by the antisymmetrization operator, because they are not the true ``identical particle'' states of the theory. This could potentially lead to new physical phenomena, which could be interpreted as departures from the Pauli Exclusion Principle (PEP). Notice that the detection of a state of the form $\ket{\xi}_P\otimes\ket{\xi}_P$ would not necessarily imply that the PEP,  and all of its physical consequences, is violated in our theory: for example, it might well be the case that the theory keeps forbidding more than two electrons  to share the same atomic orbital, in view of the aforementioned existence of classes of states that are excluded by the antisymmetrization operator.  At any rate, considering that there now are stringent bounds on PEP violations~\cite{Shi:2018wsc,PhysRevC.81.034317,Bernabei:2009zzb}, it would be interesting to study the basic physical processes that are tested by these experiments, within the context of our theory, and investigate possible observable consequences of noncommutativity. Notice that such modelization would require significant further development of the theory (at the very least, interacting QFTs with Dirac fields). Results obtained through rudimentary/simplistic methodologies, especially if they compromise Poincaré invariance, fall short of the necessary rigor and hold no significance within the context of our theoretical framework.

\section{Conclusions}

Building on our past results~\cite{Lizzi:2021rlb} and~\cite{DiLuca:2022idu}, we were able to define a QFT on the $\kappa$-Minkowski noncommutative spacetime in the same spirit as~\cite{Fiore:2007vg,Fiore:2008ta}: the coordinates of N different points cannot belong to the simple tensor product algebra, otherwise it would not be $\kappa$-Poincar\'e covariant. One needs to introduce  a \textit{braiding,} which requires a quantum R matrix for the $\kappa$-Poincar\'e group. This exists only in the lightlike case, \textit{i.e.} when the commutators between the coordinates~\eqref{eq:kmink} are described by a vector $v^\mu$ that is lightlike, or null, with respect to the metric $g_{\mu\nu}$ that is preserved by the $\kappa$-Poincar\'e group~\eqref{eq:galgebra}. Within this framework, one can define consistently covariant N-point functions, which are the backbone of QFT. The striking advantage of the approach of~\cite{Fiore:2007vg,Fiore:2008ta}, which is shared by our model, as proven in~\cite{Lizzi:2021rlb}, is that the translation-invariant combinations of different coordinates (\textit{i.e.} the coordinate differences) are commutative, which implies that the N-point functions are all commutative. This hugely simplifies the physical interpretation of the theory, as we do not have to deal with noncommutative correlation functions, whose meaning would be rather obscure. A similar conceptual simplification is achieved in many other approaches to noncommutative QFT by using a star product, and defining a path integral over commutative functions in which the action is turned into a nonlocal, infinite-derivative functional of the fields. Then, the correlation functions are commutative objects, simply obtained as expectation values or functional variations of the partition function. However, there is no sense in which these commutative N-point functions can be invariant under the quantum group of isometries of the noncommutative spacetime they are supposed to live in. In our approach, we have a way of writing commutative N-point functions which are $\kappa$-Poincar\'e invariant, and we believe that this is a key advantage of the approach based on braiding. As shown in~\cite{Fiore:2007vg,Fiore:2008ta} in the case of $\theta$-Moyal noncommutative spacetimes, with the braided structures one can define a QFT built upon the Wightman axioms, and the quantization of a free complex scalar field can be performed with the introduction of a covariant Pauli--Jordan function. In the case of $\theta$-Moyal spacetimes, the QFTs thus defined turned out to be completely indistinguishable from their commutative counterparts, as all the N-point functions of the free theory, as well as the perturbative expansion of the N-point functions of an interacting theory, turn out to be undeformed. In our lightlike $\kappa$-Minkowski case, we find that, although the Pauli-Jordan and two-point functions are undeformed, a dependence on the deformation parameter appears at the level of multiparticle states already in the free theory. The momentum, boost and charge conjugation operators are undeformed, however the creation and annihilation operators can be written, in a key advancement obtained in this paper for the first time, as an infinite nonlinear combination of undeformed creation and annihilation operators. The deformed creation operators act in a trivial way on the vacuum, and the one-particle sector looks undeformed. However, as soon as we create more than one particle we start seeing a dependence on the noncommutativity parameter: the momentum of two particles is a nonlinear combination of the two single-particle momenta. The way that two particle momenta boost under Lorentz transformations is nonlinear and mixes the momenta of the two particles. We can introduce a covariant and involutive flip operator, which acts nonlinearly on the momenta of the two particles, changing them in a more complicated way than simply exchanging them. This flip is used to define two covariant and idempotent symmetrization and antisimmetrization operators, whose image is the Fock space of bosonic and, respectively, fermionic fields. The situation is substantially simpler compared to the attempts at defining a QFT on the \textit{timelike} $\kappa$-Minkowski spacetime: in this case, the absence of a quantum R matrix makes it impossible to define a flip operator that is both involutive and Lorentz-covariant~\cite{Arzano:2008bt,Arzano:2009wp,Govindarajan:2009wt,Arzano:2013sta,Arzano:2022vmh}, which implies that the notion of identical particles and (anti-)symmetrized multiparticle states loses meaning~\cite{Arzano:2022vmh}.
We proved that our theory is C-, PT- and CPT-invariant, however P and T symmetries do not hold separately. This can already be seen at the level of the coordinate commutation relations, which break P and T symmetry.
The theory allows for the existence of states which are excluded by the Pauli principle in the classical setting. This opens up the interesting phenomenological opportunity of setting bounds on the model by experimental results searching for evidence of transitions into such states.

The noncommutative QFT defined in~\cite{Lizzi:2021rlb,DiLuca:2022idu} and completed here seems in healthy shape, and motivates interest in several future research directions. The simplest one is to write the N-point functions of the free theory for N larger than two, to check whether they are undeformed too, or perhaps the nontriviality of the multiparticle sector manifests into a dependence of higher correlators on the noncommutativity parameter. A further issue to consider is that the model studied so far is in 1+1 spacetime dimensions, and its generalization to 3+1 dimensions seems straightforward, but it has not been worked out explicitly and might yet hide some surprises. The next natural step is to introduce an interaction, which is where the theory has the highest chances of providing some predictions that depart from standard QFT on commutative Minkowski space. Further down the road, gauge theories and fermions might be explored, and perhaps a possible connection with CP violation in the Standard Model.

\section*{Acknowledgements}

G.F. acknowledges financial support by the Programme STAR Plus, funded by Federico II University and
Compagnia di San Paolo, and by the MIUR, PRIN 2017 grant 20179ZF5KS.
F.M. acknowledges support by the Q-CAYLE Proyect funded by the Regional Government of Castilla y Le\'on (Spain) and by the Ministry of Science and Innovation MICIN through NextGenerationEU (PRTR C17.I1).

\bibliographystyle{utphys_short}

\setlength\columnsep{12pt}

\begin{multicols}{2}

\end{multicols}

\end{document}